\documentclass[preprint,12pt]{elsarticle}

\usepackage{amssymb}
\usepackage{amsmath}

\usepackage{lineno}

\usepackage{textcomp}
\usepackage{stfloats}
\usepackage{url}
\usepackage{verbatim}
\usepackage{graphicx}
\usepackage{amsmath,amssymb,amsfonts,bm}
\usepackage{xcolor}
\usepackage{dsfont}
\usepackage{multirow}
\usepackage{booktabs}
\usepackage{tabularx}
\usepackage{subfig}
\usepackage{soul}

\journal{Knowledge-Based Systems}

\begin{document}

\begin{frontmatter}


\title{SinFormer: A Tailored  Transformer for Robust Radio Frequency Fingerprint Identification}


\author{Liu Yang} 
\author{Qiang Li\corref{cor1}}
\ead{lq@uestc.edu.cn}
\author{Xiaoyang Ren}

\address{School of Information and Communication Engineering, University of Electronic
	Science and Technology of China, Chengdu, China, 611731}
	
	
\cortext[cor1]{Corresponding author}


\begin{abstract}
With the rapid proliferation of wireless and Internet of Things (IoT) devices, ensuring secure and reliable device identification has become a significant challenge. Traditional security techniques, such as IP or MAC address-based authentication, are susceptible to spoofing, whereas Radio Frequency Fingerprint Identification (RFFI) offers a more secure alternative by exploiting the unique hardware imperfections in devices' RF signals. In this paper, we propose a novel deep learning-based framework for RFFI that enhances both accuracy and reliability in challenging RF environments. The core of our approach is the Signal Inception Transformer (SinFormer), which leverages a specialized multi-scale self-attention mechanism to effectively capture both large-scale and fine-grained fingerprints in signals, significantly improving identification accuracy. To further enhance robustness and reliability, we introduce a two-stage training strategy that enables the model to learn general signal features and maintain performance under adverse conditions, such as low Signal-to-Noise Ratio (SNR) or channel variations. The effectiveness of the proposed method is validated using a real-world dataset. Experimental results show that the SinFormer framework consistently outperforms existing methods in accuracy and robustness across diverse and challenging scenarios.

\end{abstract}

%
%
%


\end{frontmatter}


\section{Introduction} \label{sec:intro}
The rapid proliferation of wireless and Internet of Things (IoT) devices has made secure and reliable device identification a critical challenge. Traditional security mechanisms rely on software-based identifiers, such as Internet Protocol (IP) and Media Access Control (MAC) addresses. However, these protocol-based identities are vulnerable to various attacks~\cite{zou2016survey}. To enhance security, Radio Frequency Fingerprint Identification (RFFI) has emerged as a promising solution, offering secure and efficient device identification based on the unique hardware-level characteristics of RF signals.

\begin{figure}[]
	\centerline{\includegraphics[scale=0.55]{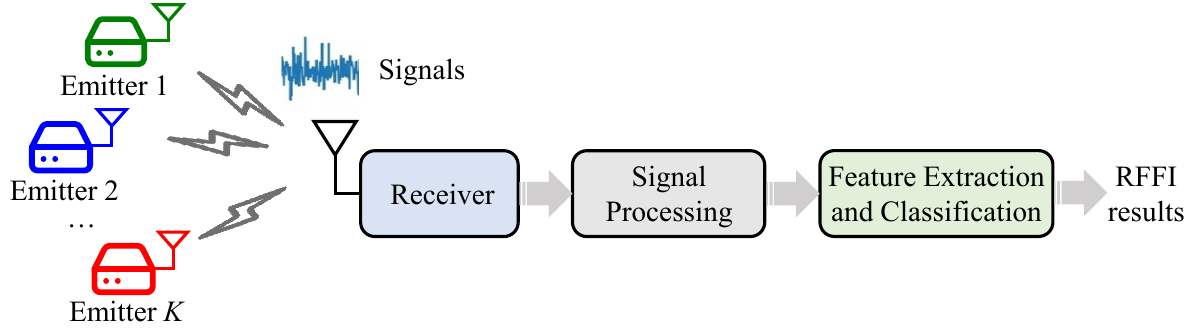}}
	\caption{
		The block diagram of the RFFI system begins with signal transmission from various emitters, represented by green, blue, and red antenna symbols. These signals are received by a receiver, where they undergo signal processing, including filtering, amplification, and analog-to-digital conversion. Following this, the feature extraction and classification stage is applied to capture and analyze unique signal characteristics or ``fingerprints'' for identification purposes.}
	\label{fig:system}
\end{figure}

\begin{figure}[]
	\centerline{\includegraphics[scale=0.65]{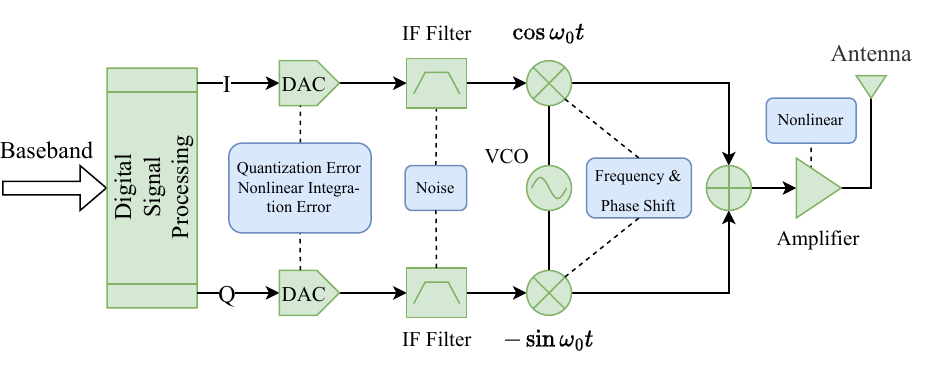}}
	\caption{The diagram of radio frequency circuit. The fingerprint of an emitter are mainly from non-ideal characteristics of individual modules.}
	\label{fig:fingerprint}
\end{figure}

RFFI identifies devices by leveraging the intrinsic physical-layer features of their RF signals, often referred to as “fingerprints”. These fingerprints, rooted in hardware imperfections, provide a robust basis for device identification, as they are difficult to forge. Figure~\ref{fig:system} depicts the RFFI system workflow. Signals transmitted by multiple emitters are received and preprocessed to enhance RF fingerprint features. Subsequently, feature extraction and classification stages leverage these features to uniquely identify devices.
Historically, RFFI has relied on manually engineered features such as Carrier Frequency Offset (CFO)~\cite{peng2018design, ma2023white, liu2019real}, In-phase/Quadrature (I/Q) imbalance~\cite{brik2008wireless, zhang2021radio}, power amplifier characteristics~\cite{polak2011identifying}, and channel state information (CSI)~\cite{fu2023deep, kong2023physical}. These hand-crafted features, illustrated in Figure~\ref{fig:fingerprint}, are used as inputs to machine learning models for classification. However, these approaches face significant limitations. Manual feature extraction is labor-intensive, relies heavily on expert knowledge, and often struggles to adapt to complex and dynamic electromagnetic environments. Consequently, such methods may fail to capture intricate RF signal patterns, resulting in reduced accuracy and robustness~\cite{xie2023disentangled, shen2021radio}.

Recently, Deep Learning (DL) has revolutionized fields like Natural Language Processing (NLP)\cite{achiam2023gpt}, Computer Vision (CV)\cite{croitoru2023diffusion}, and time series analysis~\cite{mohammadi2024deep}, offering end-to-end solutions that reduce the need for manual feature engineering. This paradigm shift has inspired researchers to explore various off-the-shelf models for RFFI. For instance, Zeng \textit{et al.}~\cite{zeng2023multi} leveraged multi-channel neural features and ResNeXt~\cite{xie2017aggregated} for device identification, while Shen \textit{et al.}~\cite{shen2021radio} utilized spectrograms and Convolutional Neural Networks (CNNs). Similarly, Ling \textit{et al.}~\cite{ling2024rsbu} employed Long Short-Term Memory (LSTM) networks to extract temporal features from RF signals. More recently, the success of Transformers~\cite{vaswani2017attention} in NLP has motivated their application in RFFI, given their ability to capture long-term dependencies and model sequential data effectively~\cite{bo2023specific}. 
However, these off-the-shelf models were originally designed for NLP or CV tasks and are not optimized for the unique challenges of RF signals. Consequently, their robustness in RFFI applications is limited, with performance degrading under conditions such as channel variations~\cite{shen2022towards, yang2023led}, noise distortions~\cite{shen2023towards}, and receiver inconsistencies~\cite{yang2024mitigating, zhao2023gan}. While strategies such as denoising and signal augmentation~\cite{wang2024design}, pre-training~\cite{huang2022deep, liu2023overcoming}, as well as domain adaptation techniques~\cite{yang2024mitigating}, have shown promise in addressing specific environmental challenges, they fall short of offering a comprehensive solution. The fundamental issue remains the lack of generalized and transferable features that can robustly handle the diverse challenges encountered in RFFI applications.

\begin{figure}[]
	\centerline{\includegraphics[scale=0.65]{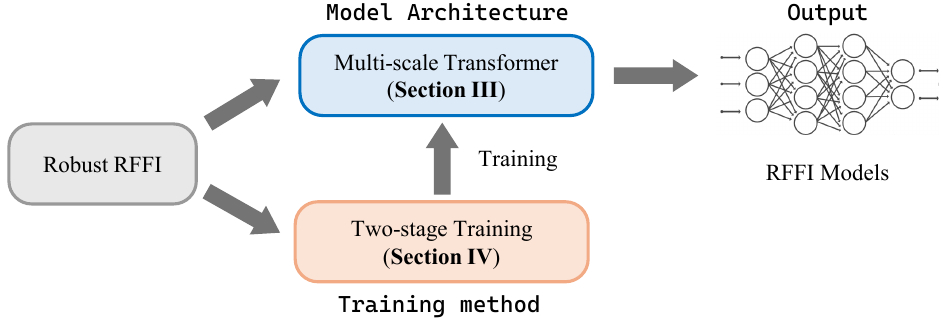}}
	\caption{
		Overview of the proposed robust RFFI framework, integrating the proposed multi-scale Transformer architecture, called SinFormer and the two-stage training method to achieve robust and adaptable RFFI models.}
	\label{fig:target}
\end{figure}

To overcome the limitations of current RFFI approaches, this paper presents a robust DL model specifically tailored for RFFI applications. Figure~\ref{fig:target} illustrates the overall structure of our proposed framework, which is built around two key components: a novel model architecture and a specialized training methodology.
First, recognizing that RF fingerprints span multiple signal-level scales, we propose the {\it Signal Inception Transformer (SinFormer)}. This model leverages a multi-scale self-attention module to effectively capture the diverse and intrinsic patterns inherent in RF signals, enhancing the model’s ability to handle complex, dynamic environments. Second, we introduce a two-stage training framework, including 1) General Feature Learning Stage: A novel unsupervised feature learning strategy is employed to extract generalized and robust features from RF signals, addressing the challenge of limited labeled data. 2) Task-Specific Adaptation Stage: The model is fine-tuned to adapt to the RFFI task, ensuring optimal identification performance across various scenarios.
The seamless integration of the SinFormer architecture with this two-stage training framework creates a unified solution capable of addressing the critical challenges in robust RFFI, such as channel variations, noise distortions, and receiver changes.

Our contributions are summarized as follows:  
\begin{itemize}
	\item Novel Model Architecture: The SinFormer model introduces a multi-scale self-attention mechanism tailored to RF signal fingerprints, enabling superior adaptability in dynamic scenarios.  
	\item Two-Stage Training Framework: The proposed training method ensures robust feature extraction and adaptability, addressing the limitations of traditional and DL-based RFFI approaches.  
	\item Benchmark Results: Comprehensive experiments validate the effectiveness of the proposed framework, establishing it as a robust and high-performing solution for future RFFI systems.  
\end{itemize}

The rest of the paper is organized as follows: 
Section~\ref{sec:model} presents system model and the RFFI problem. Section~\ref{sec:sinformer} elaborates on the proposed SinFormer model. Section~\ref{sec:pretrain} describes the  two-stage training framework. Section~\ref{sec:experiment} presents the detailed experimental results on real-world data, and Section~\ref{sec:conclusion} concludes the paper.

\section{System Model and Problem Formulation} \label{sec:model}

Let us start from the signal received model. During the $\ell$-th time interval, the received RF signal, denoted as $x_\ell(t)$, is expressed as
\begin{equation} \label{eq:system_model}
	x_\ell(t) = c(t) *  \varphi \left(s(t)\cos(\omega_c t+\theta)\right)+ n(t),
\end{equation}
where $(\ell-1)T \leq t \leq \ell T$. In this model, $n(t)$ represents the noise, $T$ is the length of the time interval, $\omega_c$ is the carrier frequency, $\theta$ is the initial phase, $s(t)$ is the random modulating signal, $c(t)$ denotes the channel response, and $*$ is the convolution operator. The function $\varphi(\cdot)$ models the nonlinear distortion induced by the emitter's hardware, which essentially encodes the unique ``fingerprint'' of the emitter. This ``fingerprint'' is extracted from $x_\ell(t)$ to distinguish the individual emitters. However, this process is influenced by noise $n(t)$ and the channel response $c(t)$, which requires the RFFI model to be robust.

\begin{figure}[]
	\centerline{\includegraphics[scale=0.45]{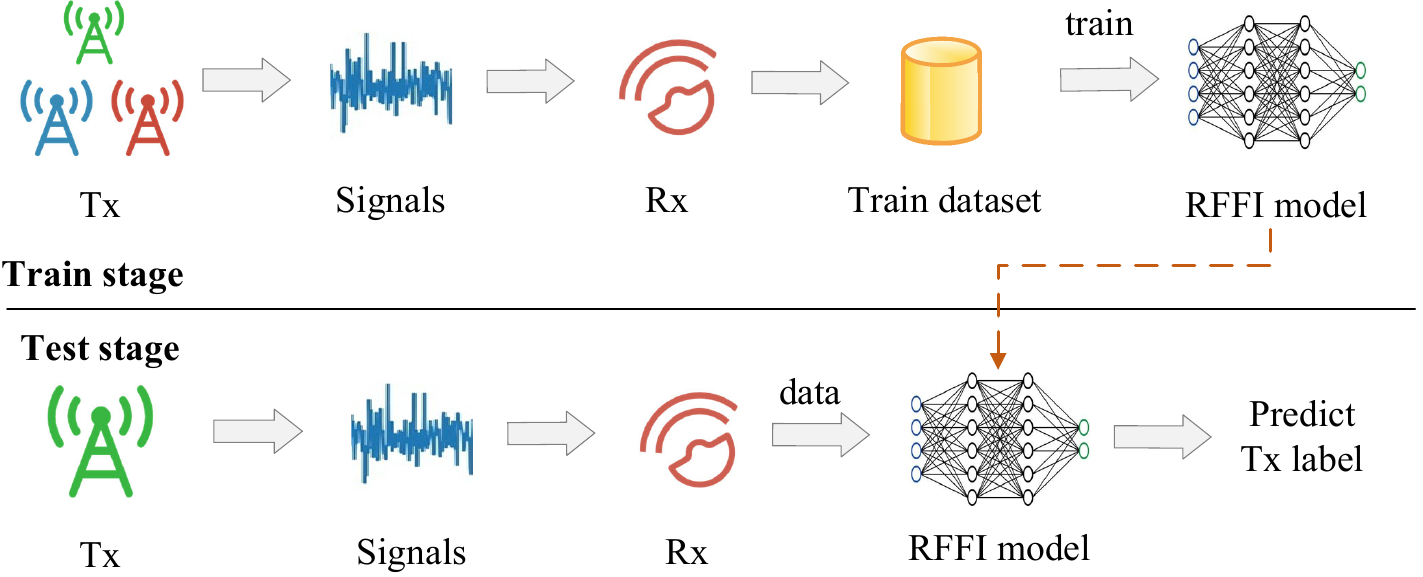}}
	\caption{The scenario of RFFI consists of a training stage and a testing stage.}
	\label{fig:scen}
\end{figure}

The RFFI process, depicted in Figure~\ref{fig:scen}, is divided into two stages: the training stage and the testing stage. 
In the training stage, signals \(x_\ell(t)\) transmitted by multiple emitters at different time intervals are collected by the receiver. Each signal is assigned an emitter label \(y_\ell \in {\cal K} \triangleq \{ 1, \ldots, K \}\), where \(K\) denotes the total number of emitters. For simplicity, we represent \(x_\ell\) as the \(\ell\)-th signal sample after pre-processing. These signal samples, along with their corresponding labels, constitute the training dataset \(\mathcal{D} = \{(x_1, y_1), \ldots, (x_N, y_N)\}\), where \(N\) is the number of samples in the dataset.
The objective is to train an RFFI model (or function) \(\mathsf{h}: {\cal X} \rightarrow {\cal K}\) using the dataset \(\mathcal{D}\) to minimize the expected classification error on the test data. Formally, this is expressed as:
\begin{equation} \label{eq:problem}
	\min_{\sf h}\mathbb{E}_{(X,Y)\sim {\cal D}} [ \mathbb{I}({\sf h}(X) \neq Y)].
\end{equation}
Here, \(\mathbb{I}(\cdot)\) denotes the indicator function, which equals \(1\) if its argument is true and \(0\) otherwise. With a slight abuse of notation, we also use \({\cal D}\) to denote the joint distribution of the data and their labels during the test stage.

\section{Signal Inception Transformer} \label{sec:sinformer}

\begin{figure} \centering 
	\subfloat[Impacts of CFO]{\includegraphics[width=1\columnwidth]{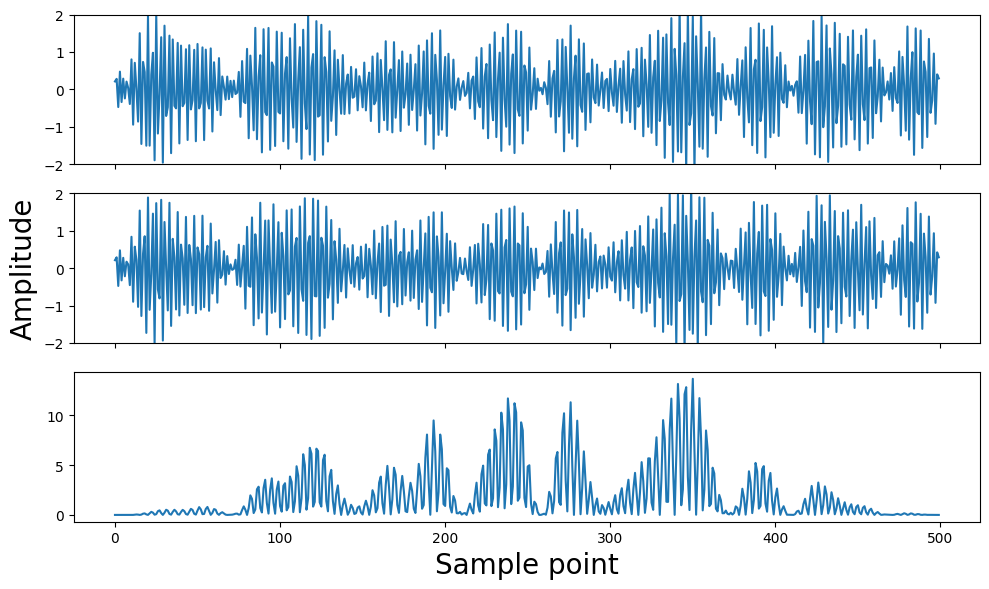}
		\label{fig:CFO}} \quad
	\subfloat[Impacts of amplifier nonlinear effects]{\includegraphics[width=1\columnwidth]{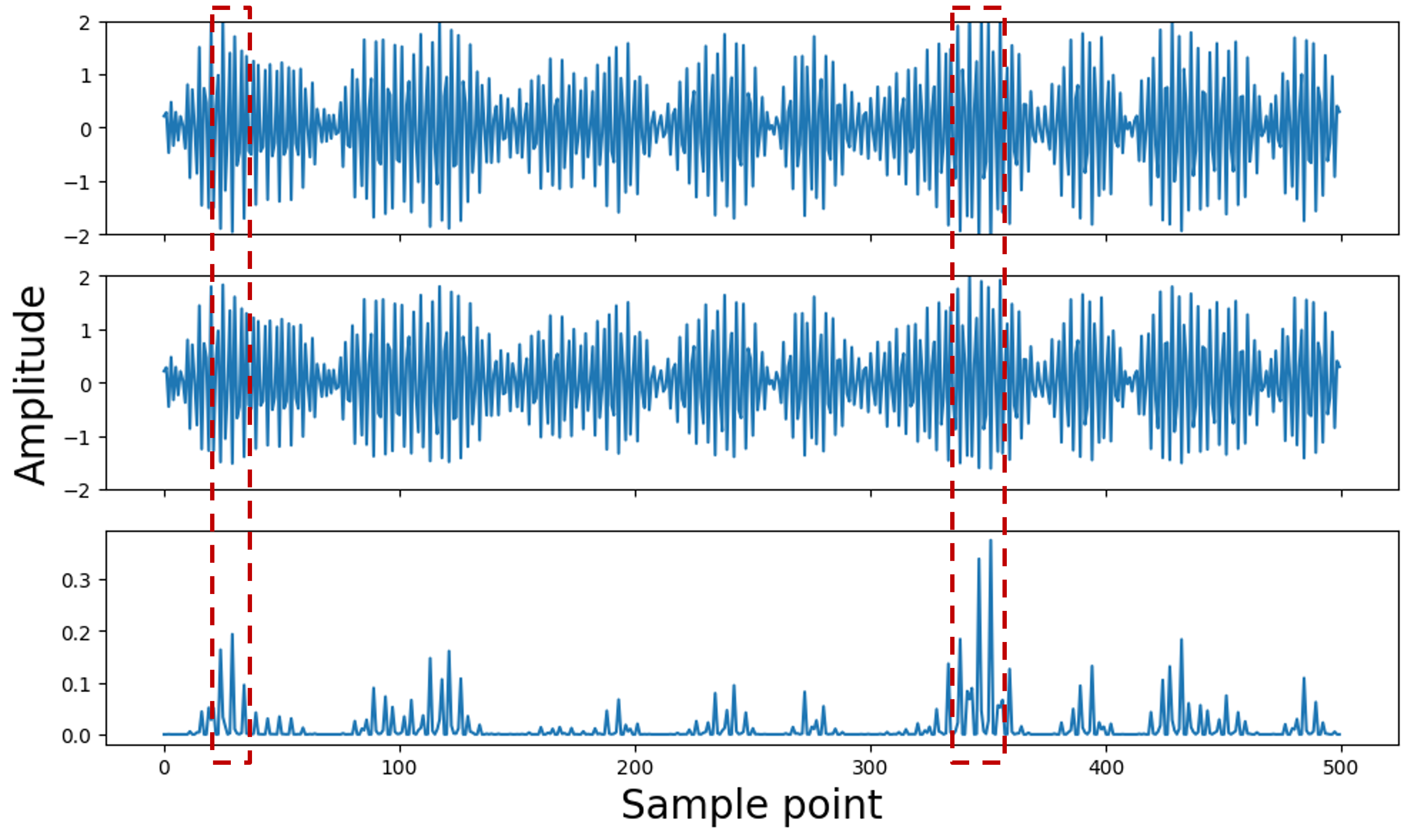}
		\label{fig:unlinear}}
	\caption{Figure of the impacts of RF fingerprint features (a) CFO, (b) amplifier nonlinear effects on signals. In each subplot, the \textbf{top} and \textbf{middle} sections show the time-domain signals after being influenced by different RF fingerprint features on the same signal. The \textbf{bottom} section represents the squared differences of the signal amplitudes at corresponding sampling points, highlighting the time-domain discrepancies between the two signals.}
	\label{fig:feature}
\end{figure}

This section aims to develop a learning model architecture specifically designed for the RFFI application. We begin by outlining the motivation and overall architecture of the proposed model, SinFormer, and then provide a detailed implementation with particular emphasis on the Inception Self-attention (ISA) module.

\subsection{Motivation and Model Overview}

The cornerstone of our design is multi-scale feature extraction. To illustrate this, we use examples of CFO and amplifier nonlinear effects to demonstrate that RF fingerprints manifest at various signal scales.

\begin{itemize}
	\item \textbf{Carrier frequency offset}: as shown in Figure~\ref{fig:feature}(a), the top and middle plots represent the time-domain signals affected by different CFO. The bottom plot displays the squared amplitude error corresponding to the sample points of the two signals. It can be observed that the impact of CFO spans the entire time domain and exhibits a periodic pattern. Additionally, the more sample points considered, the more accurately the CFO can be estimated. Therefore, CFO can be regarded as a large-scale feature and a global distortion, affecting the overall characteristics of the signal.
	
	\item \textbf{Amplifier nonlinear effects}: in Figure~\ref{fig:feature}(b), the top and middle plots show the time-domain signals affected by amplifier nonlinear effects. The bottom plot displays the squared amplitude error corresponding to the sample points of these signals. Unlike CFO, the amplifier's nonlinearity affects specific areas of the signal, causing changes only in certain parts. These small-scale features lead to sudden changes in amplitude within the red dashed boxes, showing the nonlinear behavior of the amplifier. This local impact means that while the overall signal shape stays mostly the same, the fine details change a lot. Therefore, amplifier nonlinear effects is seen as a small-scale feature and a local distortion, affecting the precise details of the signal in regions with larger amplitude.
\end{itemize}

\begin{figure}[t]
	\centerline{\includegraphics[width=1\columnwidth]{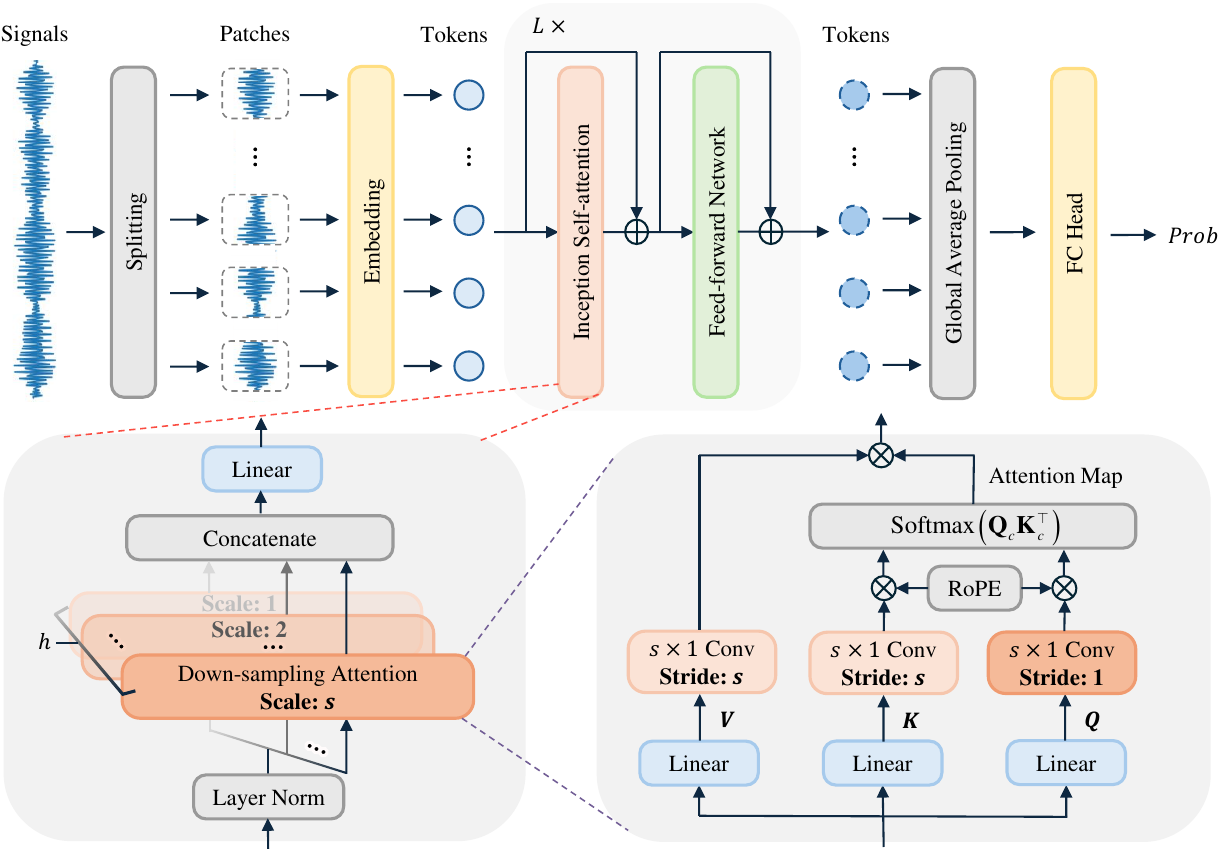}}
	\caption{The architecture of proposed SinFormer for RFFI. \textbf{Top} is the overall flow from the input signal to the output emitter class probabilities. \textbf{Bottom-left} focuses on the Inception Self-attention module, which enables both large-scale and small-scale feature interaction.  \textbf{Bottom-right} depicts the Down-sampling Attention mechanism, which facilitates targeted scale-specific feature interaction.}
	\label{fig:overview}
\end{figure}

In summary, RF signals exhibit both large-scale features, which impact the signal globally, and small-scale features, which introduce localized distortions. To accurately capture and leverage these multi-scale features, it is necessary to introduce multi-scale feature extraction capabilities into the model.
To effectively capture both large-scale and small-scale features, we propose a model SinFormer, which can integrates multi-scale feature extraction and accommodate the diverse nature of features in RF signals through an innovative architecture.

Figure~\ref{fig:overview} illustrates the architecture of the SinFormer, which is divided into three main parts: signal embedding, feature extraction, and classification. In the signal embedding module, the raw signal is first segmented into multiple patches and each patch is then transformed into a feature representation through an embedding layer. Subsequently, in the feature extraction module, these features are subsequently processed by a series of $L$ blocks, each consisting of Inception Self-Attention and Feed-forward Networks (FFN). This process allows the model to capture both large-scale and small-scale features, effectively identifying coarse-grained and fine-grained dependencies within the signal. Finally, in the classification module, the aggregated features are passed through a fully connected classification head to produce the classification probabilities for each emitter. The details of the design and implementation are provided in the following subsections.

\subsection{Signal Embedding}

The raw time-domain signal is used as the input for the SinFormer framework because it preserves RF fingerprints and natural sequence correlations. This choice aligns with the Transformer's strength in modeling dependencies, allowing SinFormer to effectively capture intricate signal relationships.  
Let $x \in \mathbb{R}^m$ be the raw signal from dataset $\mathcal{D}$, and split it into $n$ non-overlapping patches, each of length $l$, where $m = n \times l$ and $m$ is the number of sample points of raw signal. These signal patches are stacked to form a matrix ${\bf P} \in \mathbb{R}^{n \times l}$, which allows the model to focus on localized sections of each patch, making it easier to capture fine-grained features. Afterward, each patch is transformed into a higher-dimensional feature space through an embedding layer. Specifically, each row ${\bf P}_i \in \mathbb{R}^{1 \times l}$ is mapped to an embedding feature vector ${\bf X}_i \in \mathbb{R}^{1 \times d}$, viz.,
\begin{equation}
	{\bf X} = {\bf P} {\bf W}_d,
\end{equation}
where ${\bf W}_d \in \mathbb{R}^{l \times d}$ denotes the parameter matrix of the embedding layer, and $d$ is the dimension of the embedding space.
This process results in an embedding matrix ${\bf X} \in \mathbb{R}^{n \times d}$, where each row corresponds to the embedded feature of a signal patch, also called ``token''. Therefore, the matrix ${\bf X}$ is also called token sequences. By converting the raw signal patches into embedding vectors, the model can effectively learn and capture complex patterns within the signal, which are crucial for accurate classification and downstream tasks.

\subsection{Inception Self-attention}

\begin{figure}[t]
	\centerline{\includegraphics[width=1\columnwidth]{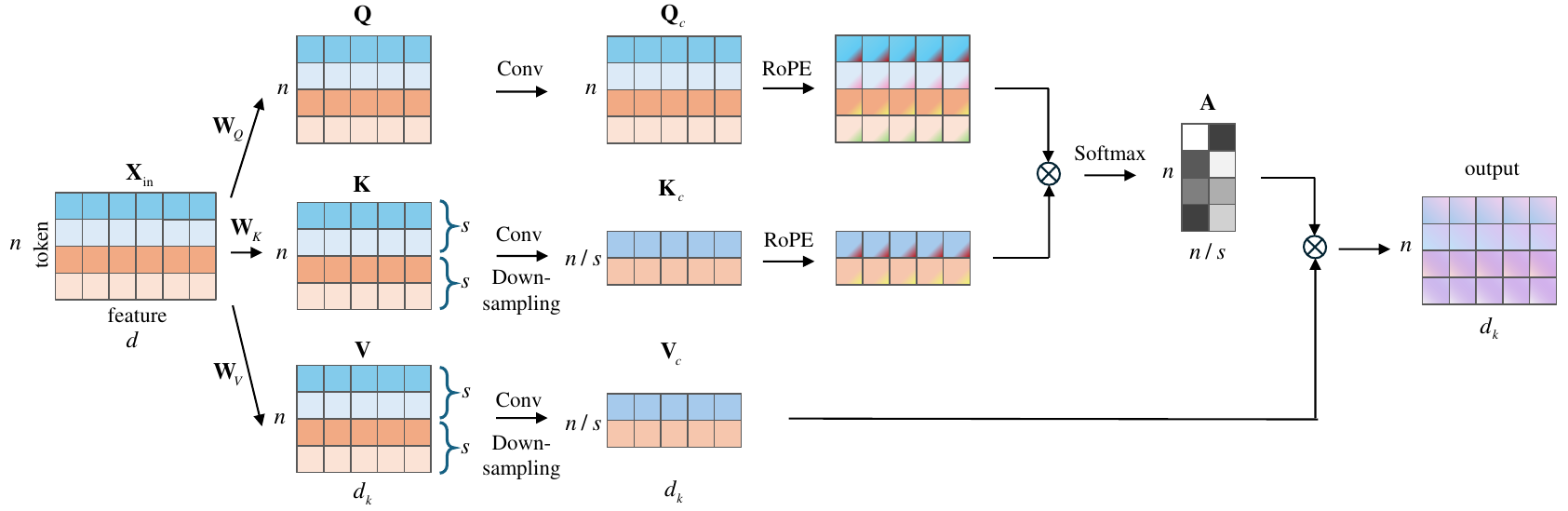}}
	\caption{The processing flow of proposed down-sampling attention module.}
	\label{fig:DSA}
\end{figure}

Although the multi-head self-attention (MHSA)~\cite{vaswani2017attention} in the vanilla Transformer is capable of modeling the dependencies and similarities between tokens, effectively mixing the features of each token, it has certain limitations. Specifically, MHSA only facilitates feature mixing between tokens on a token-to-token basis. However, the RF fingerprints exist at both large-scale and small-scale signal levels, which necessitates an attention mechanism capable of interacting with features at multiple scales. To this end, we introduce the Inception module~\cite{szegedy2015going} into MHSA to create a novel attention mechanism ISA that can effectively interact with multi-scale features.

The proposed ISA structure is illustrated in Figure~\ref{fig:overview}. Concretely, given a token sequence ${\bf X} \in \mathbb{R}^{n \times d} $, it serves as the input to $h$ parallel Down-Sampling Attention (DSA) modules after layer normalization. Each DSA module operates at a specific scale, performing feature interactions tailored to that particular scale. The outputs from all DSA modules are then concatenated along the feature dimension, creating a comprehensive multi-scale feature representation. This concatenated output is subsequently passed through a final linear projection, which can be expressed as:
\begin{equation} \label{eq:isa}
	\begin{aligned}
		\text{ISA}({\bf X}) &= \text{Concate} ( \text{scale}_1, \ldots, \text{scale}_h ) {\bf W}_O, \\ 
		\text{scale}_i &= \text{DSA}_i (\text{LN}({\bf X}); s_i),
	\end{aligned}
\end{equation}
where ${\bf W}_O \in \mathbb{R}^{hd_k \times d}$ is the parameter matrix, $d_k$ is the dimension of feature of DSA, $\text{scale}_i \in \mathbb{R}^{n \times d_k}$ is the output of $i$-th DSA, $s_i$ is the scale of $i$-th DSA, and $\text{LN}(\cdot)$ is the layer normalization operation.

For DSA with a scale factor of $s$, we apply a down-sampling approach that reduces the sequence length by a factor of $s$. This process bundles multiple tokens into groups, allowing the model to perform feature interactions at the group level rather than the individual token level. 
The processing flow of DSA is shown in Figure~\ref{fig:DSA}. 
Specifically, when the input token sequence after layer normalization ${\bf X}_{\text{in}} = \text{LN}({\bf X}) \in \mathbb{R}^{n \times d}$ is received, we first apply linear projections to obtain the query $\bf Q$, key $\bf K$, and value $\bf V$:
\begin{equation}
	{\bf Q} = {\bf X}_{\text{in}} {\bf W}_Q, {\bf K} = {\bf X}_{\text{in}} {\bf W}_K, {\bf V} = {\bf X}_{\text{in}} {\bf W}_V \in \mathbb{R}^{n \times d_k},
\end{equation}
where ${\bf W}_Q, {\bf W}_K, {\bf W}_V \in \mathbb{R}^{d \times d_k}$ are the parameter matrices.
Subsequently, we apply a convolution operation to down-sample the key and value matrices by a factor of $s$, effectively bundling every $s$ consecutive tokens into a single group, which can be represented as:
\begin{equation}
	\begin{aligned}
		{\bf K}_c &= \text{Conv}_{s,d_k,s}({\bf K}) \in \mathbb{R}^{\frac{n}{s} \times d_k}, \\
		{\bf V}_c &= \text{Conv}_{s,d_k,s}({\bf V}) \in \mathbb{R}^{\frac{n}{s} \times d_k},
	\end{aligned}
\end{equation}
where $\text{Conv}_{a_1, a_2, a_3}(\cdot)$ is one-dimensional convolution operation with kernel size $a_1$, output dimension $a_2$ and stride $a_3$. 
Next, to ensure the outputs from each DSA module in Eqn.~\eqref{eq:isa} can be seamlessly concatenated, we apply local token fusion to the query matrix without performing any down-sampling:
\begin{equation}
	{\bf Q}_c = \text{Conv}_{s,d_k,1}({\bf Q}) \in \mathbb{R}^{n \times d_k}.
\end{equation}
Then we adopt the Rotary Position Embedding (RoPE)~\cite{su2024roformer}, which encodes the absolute position with a rotation matrix and meanwhile incorporates the explicit relative position dependency in self-attention formulation. The attention map ${\bf A} \in \mathbb{R}^{n \times \frac{n}{s}}$, denotes the similarity between the $n$ mixed tokens in $\bf Q$ and the $n/s$ groups in $\bf K$, can be calculated as:
\begin{equation}
	{\bf A} = \text{Softmax} \left( \frac{\text{RoPE}({\bf Q}_c)\text{RoPE}({\bf K}_c)^\top}{\sqrt{d_k}} \right).
\end{equation}
Finally, the attention map $\bf A$ is multiplied with the value matrix to produce the output of the DSA module. This operation can be expressed as:
\begin{equation}
	\text{DSA}({\bf X}_{\text{in}}; s) = {\bf A} {\bf V}_c,
\end{equation}
which denotes each token of the output is the linear combination of tokens in ${\bf V}_c$.

In summary, ISA contains multiple DSA at different scales to facilitate feature interactions across various levels of granularity. This approach not only enables token-to-token interactions but also allows for token-to-group and group-to-group interactions. By integrating these multi-scale feature interactions, ISA effectively captures a richer and more diverse set of features, enhancing the model's ability to signal understanding.

\subsection{Feed-forward Network and Residual Connection}
Unlike ISA module, the FFN consists of two fully connected layers, which perform a series of nonlinear transformations on the input feature. 
It enables the model to learn and represent complex patterns in the data and can be represented as:
\begin{equation} \label{eq:ffn}
	\text{FFN}({\bf X}) = \phi(LN({\bf X}) {\bf W}_1 + {\bm b}_1) {\bf W}_2 + {\bm b}_2,
\end{equation}
where ${\bf W}_1 \in \mathbb{R}^{d \times d_f}$ and ${\bf W}_2 \in \mathbb{R}^{d_f \times d}$ are weights of linear projection, ${\bm b}_1 \in \mathbb{R}^{d_f}$ and ${\bm b}_2 \in \mathbb{R}^{d}$ are biases, $d_f$ denotes the dimension of the projected higher-dimensional space, $\phi(\cdot)$ is a nonlinear activation function, and in this context, we use the Gaussian Error Linear Unit (GELU).

Similar to the vanilla Transformer and ResNet~\cite{he2016deep}, the SinFormer encoder also employs residual connections to facilitate stable and effective training. The encoder can be expressed as:
\begin{equation}
	\begin{aligned}
		{\bf X}' &= \text{ISA}({\bf X}^{\ell - 1}) + {\bf X}^{\ell - 1}, \\
		{\bf X}^\ell &= \text{ISA}({\bf X}') + {\bf X}',
	\end{aligned}
\end{equation}
where ${\bf X}^{\ell - 1}$ is the output of the $(\ell - 1)$-th block.

\subsection{Classification}
The output from the encoder, denoted as ${\bf X}^L$, undergoes global average pooling. This operation averages the features across the token dimension, transforming the output into a single feature vector ${\bm z} \in \mathbb{R}^d$, where:
\begin{equation}
	{\bm z} = \frac{1}{n} \sum_{i=1}^{n} {\bf X}_i^L.
\end{equation}
Next, $z$ is fed into a fully connected (FC) head, which applies a linear transformation followed by a softmax activation function to produce the probability distribution over emitters:
\begin{equation}
	{\bm p} = \text{Softmax}({\bf W}_{\text{FC}} z + {\bm b}_{\text{FC}}),
\end{equation}
where ${\bf W}_{\text{FC}} \in \mathbb{R}^{k \times d}$ and ${\bm b}_{\text{FC}} \in \mathbb{R}^{k}$ are the weight matrix and bias of the fully connected layer, respectively, and $k$ represents the number of emitter classes.

So far, we have provided a comprehensive description of the entire SinFormer architecture. In the following section, we will address the efficient and robust training methodologies associated with SinFormer.

\section{Two-stage training} \label{sec:pretrain}

\begin{figure}[t]
	\centerline{\includegraphics[width=1\columnwidth]{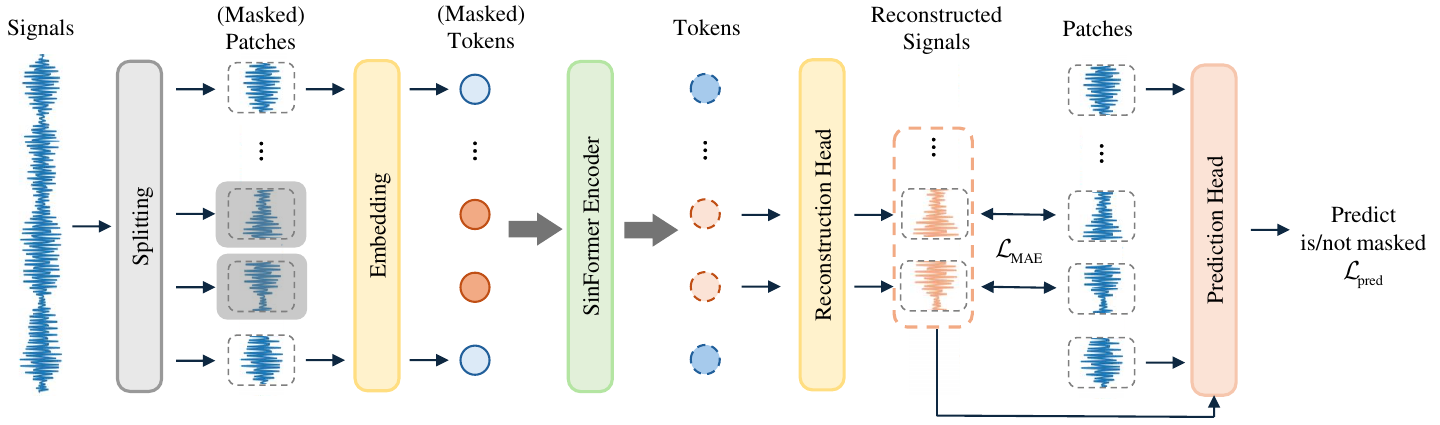}}
	\caption{The overview of the proposed unsupervised general feature learning approach, MAE-AC.}
	\label{fig:pretrain}
\end{figure}

The two-stage training process includes unsupervised general feature learning stage and supervised RFFI task-oriented adaptation stage. The first stage focuses on learning efficient fingerprint representations from RF signals in an unsupervised manner, while the second stage fine-tunes SinFormer using labeled signal-emitter data to adapt it to the RFFI task. This approach ensures the model achieves a balance between generalization and specialization, enhancing its robustness and adaptability to diverse real-world scenarios.

\subsection{Stage 1: General Feature Learning}

Inspired by pre-training paradigms in CV and NLP, we employ unsupervised pre-training with techniques like the Masked Autoencoder (MAE)~\cite{he2022masked}, where the model reconstructs masked signal patches to learn generalized features from RF signals.  These features are task-independent, enhancing the model's robustness to environmental changes such as channel variations, receiver diversity, or low SNR.  We should mention that the pre-training for RFFI  presents unique challenges. Unlike CV and NLP, where abundant and well-organized pre-training datasets are readily available online, there is no comparable large-scale, off-the-shelf RF dataset for pre-training. As a result, we pre-train and fine-tune SinFormer on the same dataset \({\cal D}\), diverging from typical CV and NLP pre-training setups. Additionally, the limited volume and diversity of RF data increase the risk of overfitting during general feature learning. To address this, we augment the standard reconstruction task with an auxiliary classification task. This auxiliary task, implemented via an additional classification head, differentiates between masked and unmasked signal patches. By combining these tasks, SinFormer learns richer and more generalized representations while reducing overfitting. Figure~\ref{fig:pretrain} provides an overview of the proposed unsupervised general feature learning approach, MAE-AC. Specifically, given an unlabeled signal, it is first divided into patches and then mapped into feature representations via embedding layer, resulting in a sequence of tokens ${\bf X} \in \mathbb{R}^{n \times d}$.

\subsubsection{Masking tokens}

After obtaining \({\bf X}\), a portion of the tokens are randomly masked and replaced with a placeholder token indicating their absence, while the unmasked tokens retain their original values: 
\[
{\bf X}_\text{mask} = \begin{cases}
	{\bf X}_i, & \text{mask}_i = 0, \\
	\mathcal{M}, & \text{mask}_i = 1,
\end{cases}
\]
where \({\bf X}_i \in \mathbb{R}^d\) is the \(i\)-th row (token) of the token sequence \({\bf X}\), \(\text{mask}_i\) is a uniform binary random variable (\(\text{mask}_i = 1\) indicates masking and \(\text{mask}_i = 0\) no masking), and \(\mathcal{M} \in \mathbb{R}^d\) is the placeholder token replacing the masked tokens. The placeholder token is randomly initialized and remains fixed throughout training, ensuring the model reconstructs the missing information based solely on the context from unmasked tokens, rather than patterns learned from the masking token itself.

\subsubsection{Reconstructing masked patches}
The masked sequence is fed into the SinFormer encoder, which processes the sequence and outputs a set of encoded features. These features are then passed through a reconstruction head, implemented as a linear layer, to transform the encoded features back into the original input signal patch space. This process is expressed as:  
\[
\hat{{\bf P}} = \text{Encoder}({\bf X}_\text{mask}) {\bf W}_R,
\]
where \(\text{Encoder}(\cdot)\) represents SinFormer's encoder, comprising a series of ISA and FFN modules. The processed features, \(\text{Encoder}({\bf X}_\text{mask}) \in \mathbb{R}^{n \times d}\), and the parameter matrix of the reconstruction head, \({\bf W}_R \in \mathbb{R}^{d \times l}\), are used to map the encoded features to the original signal patch space.

In the MAE framework, the focus is on accurately reconstructing the masked portions of the input tokens. The reconstruction loss is calculated by comparing the reconstructed patches of the masked tokens with their original patches:  
\[
\mathcal{L}_{\text{MAE}} = \frac{1}{\sum_{i=1}^{n} { \text{mask}_i }} \|\hat{{\bf P}}_i - {\bf P}_i\|_2^2,
\]
where \(\hat{{\bf P}}_i \in \mathbb{R}^l\) and \({\bf P}_i \in \mathbb{R}^l\) are the \(i\)-th rows of the patch matrices \(\hat{{\bf P}}\) and \({\bf P}\), respectively, and \(\| \cdot \|_2\) denotes the \(\ell_2\)-norm. By focusing on the masked tokens, the MAE loss encourages the model to capture the structural information of the signal, ensuring accurate recovery of the missing data.

\subsubsection{Auxiliary classification task}
While MAE effectively learns representations by reconstructing masked portions of input data, it has limitations. Specifically, MAE may cause the model to overemphasize reconstructing fine details or outliers in the signal, leading to overfitting—especially with small pre-training datasets like ours. As a result, the model may struggle to generalize to new data and environments, being overly tuned to specific training signal details.

To mitigate this, we propose an auxiliary classification task to complement reconstruction. This task focuses on capturing global trends and structures rather than just fine details. It involves distinguishing between unmasked original patches and reconstructed masked patches, encouraging the model to learn representations that generalize better to unseen signals and RFFI tasks. The auxiliary classification loss is formulated as:  
\begin{equation}
		\mathcal{L}_\text{aux} = -\frac{1}{n} \big[\sum_{i=1}^{n}  \text{mask}_i \log {\sigma (\text{FFN} (\hat{\bf P}_i))} + (1-\text{mask}_i) \log {\sigma (\text{FFN} ({\bf P}_i))}   \big],
\end{equation}
where $\sigma(\cdot)$ is the sigmoid function, which maps the output of the feed-forward network (FFN) to a probability value between 0 and 1, $\text{FFN}(\cdot)$ is the same as Eqn.~\eqref{eq:ffn}, and ${\bf W}_1 \in \mathbb{R}^{l \times d_f}, {\bf W}_2 \in \mathbb{R}^{d_f \times 1}, {\bm b}_1 \in \mathbb{R}^{d_f}, {\bm b}_2 \in \mathbb{R}$ are parameters of FFN. Then, we minimize the combined loss function used in MAE-AC, which can be expressed as:
\begin{equation}
	\mathcal{L}_\text{pretrain} = \mathcal{L}_\text{MAE} + \gamma \mathcal{L}_\text{aux},
\end{equation}
where $\gamma > 0$ is hyper-parameter. The proposed auxiliary classification task enhances the MAE framework by guiding the model to learn more robust and generalized feature representations, ultimately leading to improved performance in the following task-specific adaptation stage.

\subsection{Stage 2: Task-specific Adaptation}
In this phase, SinFormer is fine-tuned on the labeled signal dataset \(\mathcal{D}\) to better align with the RFFI task. This is achieved through supervised learning by minimizing a cross-entropy loss:  
\[
\mathcal{L}_{\text{cls}} = - \sum_{c=1}^{K} [\text{onehot}(y)]_c \log{p_c},
\]
where \([\text{onehot}(y)]_c = 1\) if \(c = y\), and \(0\) otherwise.  

To enhance training efficiency and improve model robustness, we incorporate self-supervised auxiliary tasks (SSAT)~\cite{das2024limited} in this stage. The SSAT loss is defined as:  
\[
\mathcal{L}_{\text{SSAT}} = \frac{1}{n \cdot l} \| {\bf X}^L {\bf W}_{\text{R}} - {\bf P} \|_F^2,
\]
where \({\bf W}_{\text{R}} \in \mathbb{R}^{d \times l}\) is a learnable parameter matrix, and \(\|\cdot\|_F\) is the Frobenius norm.  

The total loss for joint optimization combines the classification and auxiliary tasks:  
\[
\mathcal{L}_{\text{task}} = \alpha \mathcal{L}_{\text{cls}} + \beta \mathcal{L}_{\text{SSAT}},
\]
where \(\alpha\) and \(\beta\) are positive hyperparameters balancing the contributions of the two tasks.  

By fine-tuning all layers, this task-specific adaptation phase optimizes the model to achieve greater accuracy and robustness for RFFI. Leveraging the generalized features learned during unsupervised pre-training, the model converges faster and performs better than training from scratch.

\section{Experiment} \label{sec:experiment}

In this section, we first describe the real-world dataset and evaluate the proposed method's performance by comparing it with state-of-the-art approaches in RFFI and time series classification.   We also present ablation studies and demonstrate the method’s effectiveness across varying SNR levels, in unknown emitter recognition tasks, and under practical wireless conditions including narrowband interference, multipath propagation, and large-scale scenarios.


\subsection{Dataset description}

The data used in this experiment was collected from eight radios operating in a controlled laboratory environment. The transmitted signals are OFDM-modulated, with a total of 2192 subcarriers. Each subcarrier is modulated using QPSK. The duration of each signal segment is 700 $\mu s$, with the first 100 $\mu s$ dedicated to a pilot sequence, followed by the signal data packet.
The communication setup involves networking the radios, allowing each radio to communicate with the others. The signals from the radios are transmitted into the open environment using a power divider, with the transmit power of the target radio set to 1W, while the other radios are set to 0.001W. These radiated signals are then received by an antenna that captures the signals from different radios. The signal frequency is set to 2230 MHz.

To facilitate signal processing, the received signals are down-converted to an intermediate frequency (IF) of 77 MHz using the same frequency conversion equipment. The equipment settings are as follows: a sampling rate of 102.4 MHz, RF attenuation of 10 dB, IF attenuation of 6 dB, standard mode, and a channel bandwidth of 40 MHz.
Signal collection is performed using a single-channel sampler with a sampling rate of 200 MHz, effective 12-bit resolution, and the signals are stored in int16 format. Both the transmitting and receiving antennas were fixed throughout the experiment to ensure consistent signal capture.

The data collection process was conducted in both indoor and outdoor environments, with multiple sessions collected at different times. For the indoor scenario, data was collected in 9 consecutive sessions, named In-1 to In-9. For the outdoor scenario, signals were collected in 2 sessions, labeled Out-1 and Out-2. After collecting the signals, we applied energy detection to extract the relevant signal portions and segmented them into signal samples, each containing 2000 sample points. Each session includes 50,000 signal samples per transmitter.
To prepare the dataset for training and evaluation, we randomly assigned 40,000 signal samples from each session and each emitter to the training set, with the remaining 10,000 signal samples designated for the test set. 

\begin{table}[t] \scriptsize
	\centering
	 
		\belowrulesep=0pt
		\aboverulesep=0pt
		\renewcommand\arraystretch{1.3}
		\caption{\label{table:model} The details of the SinFormer model}
		\begin{tabular}{c | c | c | c c}
			\toprule
			
			\textbf{Stage} & \textbf{Structure} & \textbf{Parameters} & \textbf{Values}  \\
			\hline
			\multirow{2}{*}{Embedding} & \multirow{2}{*}{Linear} & Input dimension & 100  \\
			\cline{3-4}
			~ & ~ & Output dimension & 256 \\
			\hline
			\multirow{6}{*}{Encoder block $\times$ 6} & \multirow{3}{*}{ISA module} & The number of DSA module & 16  \\
			\cline{3-4}
			~ & ~ & Scale & [1, 2, 5, 10] $\times$ 4  \\
			\cline{3-4}
			~ & ~ & $\mathbf{Q}_c,\mathbf{K}_c,\mathbf{V}_c$ dimension & 64  \\
			\cline{2-4}
			~ & \multirow{3}{*}{FFN} & Input dimension & 256  \\
			\cline{3-4}
			~ & ~ & Hidden dimension & 512  \\
			\cline{3-4}
			~ & ~ & Output dimension & 256  \\
			\hline
			\multirow{2}{*}{FC head} & \multirow{2}{*}{Linear} & Input dimension & 256  \\
			\cline{3-4}
			~ & ~ & Output dimension & 8 \\
			\bottomrule
	\end{tabular}
\end{table}

\begin{table}[t] \scriptsize
	\belowrulesep=0pt
	\aboverulesep=0pt
	\renewcommand\arraystretch{1.3}
	\centering
	\caption{\label{table:complexity} The computational efficiency of the models}
	\begin{tabular}{c | c c}
		\toprule
		
		\textbf{Method} & \textbf{No. of parameters} & \textbf{FLOPs}   \\
		\hline
		GRU-GCN & 6.750 M & 157.929 M \\
		ResNext & 8.273 M & 54.88 M \\
		ViT & 7.906 M & 166.098 M \\
		PatchMixer & 8.026 M & 90.595 M \\
		GLFormer & 6.369 M & 127.197 M \\
		PVT & 10.646 M & 35.343 M \\
		DeiT & 7.908 M & 173.990 M \\
		Vision Mamba & 6.794 M & 48.600 M \\
		SinFormer (proposed) & 7.942 M & 158.618 M \\
		\hline
		
		\bottomrule
	\end{tabular}
\end{table}

\begin{table}[t] \scriptsize
	\belowrulesep=0pt
	\aboverulesep=0pt
	\renewcommand\arraystretch{1.3}
	\centering
	\caption{\label{table:hyperparameters} Hyper-parameters of Experiments}
	\begin{tabular}{c | c}
		\toprule
		\textbf{Hyper-parameters} & \textbf{Values}  \\
		\hline
		Patch length $l$ & 100 \\
		\hline
		Trade-off hyper-parameters $\alpha, \beta, \gamma$ & 0.2, 0.8, 0.2\\
		\hline
		Optimizer & Adam \\
		\hline
		Epoch & 50 \\
		\hline
		Batch size & 1024 \\
		\hline
		Learning rate & 0.0006 \\
		
		\bottomrule
	\end{tabular}
\end{table}

\subsection{Model setting}
To ensure reproducibility and clarify our design choices, we detail the key hyperparameter settings used in the proposed SinFormer model. As shown in Table~\ref{table:model}, the input embedding dimension is set to 100 based on the characteristics of the raw signal, and the output embedding dimension is set to 256 to balance feature richness and computational efficiency. Sixteen DSA modules are employed with a repeated scale pattern of [1, 2, 5, 10], chosen to evenly cover the 20 input patches for effective multi-scale feature extraction. The dimensions of the query, key, and value vectors are set to 64, which offers a good trade-off between model expressiveness and computational cost. The feed-forward network adopts a hidden dimension of 512, following common Transformer design practices to ensure sufficient non-linearity and transformation capacity. Finally, the output layer dimension is set to 8, corresponding to the number of classification categories in the RFFI task.

\subsection{Evaluation results} \label{sec:results}

We compared the proposed SinFormer (in Section~\ref{sec:sinformer}) and two-stage training (in Section~\ref{sec:pretrain}) with several different methods, including those based on RNNs, CNNs, and Transformers. For the RNN-based models, we selected the enhanced GRU model, GRU-FCN~\cite{elsayed2018deep}. In the CNN category, we employed ResNext~\cite{xie2017aggregated}, a model widely used in RFFI tasks, to benchmark performance, as well as PatchMixer~\cite{gong2023patchmixer}, an advanced CNN-based method for time series forecasting. 
  For Transformer-based baselines, we included Vision Transformer (ViT)~\cite{dosovitskiy2020image}, Pyramid Vision Transformer (PVT)~\cite{Wang_2021_ICCV}, Data-efficient image Transformers (DeiT)~\cite{pmlr-v139-touvron21a}, and Vision Mamba~\cite{liu2024mamba}, a recently proposed state-space model that achieves competitive results in various sequence modeling tasks. In addition, we adopted GLFormer~\cite{deng2023lightweight}, a lightweight Transformer specifically designed for RFFI. To further benchmark against state-of-the-art RFFI systems, we also included Adaptive Semantic Augmentation (ASA)~\cite{cai2025toward}, a representative high-performance RFFI method. The parameter counts and floating-point operations (FLOPs) of these models are summarized in Table~\ref{table:complexity}, providing a quantitative comparison of their model sizes and inference costs.


\setlength{\aboverulesep}{0pt}
\setlength{\belowrulesep}{0pt}
\begin{table}[t] \scriptsize
	\renewcommand\arraystretch{1.3}
	\centering
	\caption{\label{table:compare-to-others} Classification accuracy (\%) of comparative experiments}
	\resizebox{1\columnwidth}{!}{
		\begin{tabular}{c|c|ccccccccccc}
			\toprule
			\textbf{\begin{tabular}[c]{@{}c@{}}Number of \\training samples \\ \end{tabular}} & {\textbf{Session}} & GRU-FCN & ResNext & ViT & PatchMixer & PVT & DeiT & Vison Mamba & ASA & GLFormer & SinFormer & \begin{tabular}[c]{@{}c@{}}SinFormer \\ + two-stage training\end{tabular} \\
			\hline
			\multirow{5}{*}{40000 $\times$ 8 $\times$ 7}                                    & In-1 $\sim$ In-7                  & 95.18{\tiny$\pm$0.10}                   & 96.63{\tiny$\pm$3.02}                   & \textbf{99.71}{\tiny$\pm$0.04}               & 73.51{\tiny$\pm$0.09}                      &
			98.93{\tiny$\pm$0.14}                   & 99.37{\tiny$\pm$0.08}                   & 99.15{\tiny$\pm$0.13}               & 98.95{\tiny$\pm$0.73} & 96.08{\tiny$\pm$0.12}                    & 99.36{\tiny$\pm$0.13}                     & \underline{99.60}{\tiny$\pm$0.05}                              \\
			& In-8                                  & 62.31{\tiny$\pm$1.75}                   & 61.93{\tiny$\pm$2.53}                   & 65.91{\tiny$\pm$1.47}               & 48.39{\tiny$\pm$0.47}                      & 67.03{\tiny$\pm$0.76}                   & 64.12{\tiny$\pm$1.95}                   & 65.33{\tiny$\pm$1.17}               & 63.90{\tiny$\pm$2.36}                      & 62.74{\tiny$\pm$0.26}                    & \underline{67.16}{\tiny$\pm$1.95}                     & \textbf{69.10}{\tiny$\pm$2.09}                              \\
			& In-9                                  & 58.79{\tiny$\pm$1.19}                   & 49.60{\tiny$\pm$6.49}                   & 61.02{\tiny$\pm$1.69}               & 56.00{\tiny$\pm$0.24}                      & 62.14{\tiny$\pm$1.02}                   & 57.43{\tiny$\pm$2.12}                   & 62.63{\tiny$\pm$1.97}               & 54.25{\tiny$\pm$1.51}                      & 57.19{\tiny$\pm$0.34}                    & \textbf{63.97}{\tiny$\pm$0.98}                     & \underline{63.13}{\tiny$\pm$1.58}                              \\
			& Out-1                                 & 47.13{\tiny$\pm$0.57}                   & 36.34{\tiny$\pm$2.12}                   & 52.84{\tiny$\pm$1.06}               & 25.95{\tiny$\pm$0.46}                      & 50.20{\tiny$\pm$1.31}                   & 54.86{\tiny$\pm$0.68}                   & 52.65{\tiny$\pm$1.33}               & 48.12{\tiny$\pm$1.68}                      & 45.63{\tiny$\pm$0.79}                    & \textbf{57.08}{\tiny$\pm$0.98}                     & \underline{55.32}{\tiny$\pm$0.70}                              \\
			& Out-2                                 & 63.76{\tiny$\pm$1.51}                   & 57.47{\tiny$\pm$5.20}                   & 70.97{\tiny$\pm$1.50}               & 24.74{\tiny$\pm$0.19}                      & 73.84{\tiny$\pm$1.68}                   & 76.40{\tiny$\pm$1.44}                   & 71.49{\tiny$\pm$2.32}               & 70.40{\tiny$\pm$2.13}                      & 71.13{\tiny$\pm$1.15}                    & \textbf{82.39}{\tiny$\pm$1.74}                     & \underline{82.02}{\tiny$\pm$1.77}                              \\
			\hline
			\multirow{5}{*}{30000 $\times$ 8 $\times$ 7}                                    & In-1 $\sim$ In-7                  & 63.94{\tiny$\pm$0.13}                   & 95.44{\tiny$\pm$3.96}                   & 99.21{\tiny$\pm$0.04}               & 66.71{\tiny$\pm$0.07}                                                           &98.62{\tiny$\pm$0.12}                   & \textbf{99.54}{\tiny$\pm$0.06}                   & 98.96{\tiny$\pm$0.08}               & 98.43{\tiny$\pm$0.25}                                                           & 95.55{\tiny$\pm$0.11}                    & 99.05{\tiny$\pm$0.13}                     & \underline{99.31}{\tiny$\pm$0.10}                              \\
			& In-8                                  & 35.50{\tiny$\pm$0.89}                   & 58.77{\tiny$\pm$3.71}                   & 66.99{\tiny$\pm$1.77}               & 43.58{\tiny$\pm$0.53}                                                           &66.96{\tiny$\pm$1.27}                   & 63.73{\tiny$\pm$1.30}                   & 66.21{\tiny$\pm$1.30}               & 63.84{\tiny$\pm$1.15}                                                           & 61.77{\tiny$\pm$0.23}                    & \underline{67.22}{\tiny$\pm$1.45}                     & \textbf{69.53}{\tiny$\pm$2.09}                              \\
			& In-9                                  & 42.34{\tiny$\pm$0.39}                   & 47.61{\tiny$\pm$4.47}                   & 60.40{\tiny$\pm$1.87}               & 50.23{\tiny$\pm$0.09}                                                           &62.34{\tiny$\pm$1.27}                   & 58.13{\tiny$\pm$1.65}                   & 62.63{\tiny$\pm$0.77}               & 53.10{\tiny$\pm$1.61}                                                           & 55.94{\tiny$\pm$0.40}                    & \underline{64.16}{\tiny$\pm$1.13}                     & \textbf{65.14}{\tiny$\pm$1.81}                              \\
			& Out-1                                 & 26.70{\tiny$\pm$0.41}                   & 39.24{\tiny$\pm$2.78}                   & 54.20{\tiny$\pm$1.25}               & 26.22{\tiny$\pm$3.71}                                                           &50.44{\tiny$\pm$1.04}                   & 53.08{\tiny$\pm$1.36}                   & 54.09{\tiny$\pm$1.05}               & 47.14{\tiny$\pm$1.64}                                                           & 47.36{\tiny$\pm$0.45}                    & \underline{54.42}{\tiny$\pm$1.36}                     & \textbf{55.46}{\tiny$\pm$1.40}                              \\
			& Out-2                                 & 20.02{\tiny$\pm$0.18}                   & 56.96{\tiny$\pm$6.93}                   & 74.08{\tiny$\pm$2.00}               & 23.54{\tiny$\pm$0.10}                                                           &73.15{\tiny$\pm$1.29}                   & 74.40{\tiny$\pm$2.40}                   & 73.74{\tiny$\pm$1.02}               & 70.11{\tiny$\pm$1.25}                                                           & 70.74{\tiny$\pm$0.87}                    & \underline{80.59}{\tiny$\pm$1.25}                     & \textbf{81.70}{\tiny$\pm$2.07}                              \\
			\hline
			\multirow{5}{*}{20000 $\times$ 8 $\times$ 7}                                    & In-1 $\sim$ In-7                  & 48.74{\tiny$\pm$0.07}                   & 95.34{\tiny$\pm$3.81}                   & \textbf{98.15}{\tiny$\pm$0.15}               & 46.58{\tiny$\pm$0.09}                      &96.56{\tiny$\pm$0.21}                   & \underline{97.78}{\tiny$\pm$0.08}                   & 92.01{\tiny$\pm$0.35}               & 94.52{\tiny$\pm$0.42}                      & 93.79{\tiny$\pm$0.21}                    & 96.33{\tiny$\pm$0.22}                     & 97.38{\tiny$\pm$0.08}                              \\
			& In-8                                  & 26.64{\tiny$\pm$0.75}                   & 59.51{\tiny$\pm$1.85}                   & 65.12{\tiny$\pm$1.53}               & 28.11{\tiny$\pm$0.33}                      & 66.86{\tiny$\pm$1.31}                   & 63.57{\tiny$\pm$1.19}                   & 59.18{\tiny$\pm$1.19}               & 60.08{\tiny$\pm$1.40}                      & 63.91{\tiny$\pm$0.77}                    & \underline{69.58}{\tiny$\pm$0.89}                     & \textbf{70.07}{\tiny$\pm$0.83}                              \\
			& In-9                                  & 32.31{\tiny$\pm$0.35}                   & 47.53{\tiny$\pm$6.13}                   & \underline{62.75}{\tiny$\pm$1.43}               & 34.57{\tiny$\pm$0.19}                      &62.16{\tiny$\pm$1.39}                   & 60.73{\tiny$\pm$1.29}                   & 58.16{\tiny$\pm$1.03}               & 52.86{\tiny$\pm$1.20}                      & 56.61{\tiny$\pm$0.36}                    & 61.18{\tiny$\pm$1.65}                     & \textbf{62.88}{\tiny$\pm$1.69}                              \\
			& Out-1                                 & 24.35{\tiny$\pm$0.39}                   & 40.42{\tiny$\pm$2.98}                   & 51.26{\tiny$\pm$1.11}               & 24.02{\tiny$\pm$0.17}                      & 48.27{\tiny$\pm$0.70}                   & 51.57{\tiny$\pm$1.29}                   & 50.55{\tiny$\pm$0.72}               & 47.51{\tiny$\pm$1.21}                      & 46.22{\tiny$\pm$0.52}                    & \underline{52.69}{\tiny$\pm$1.36}                     & \textbf{53.91}{\tiny$\pm$1.63}                              \\
			& Out-2                                 & 17.10{\tiny$\pm$0.20}                   & 60.76{\tiny$\pm$7.60}                   & 70.04{\tiny$\pm$1.75}               & 18.33{\tiny$\pm$0.08}                      & 70.41{\tiny$\pm$1.70}                   & 71.42{\tiny$\pm$1.71}                   & 61.73{\tiny$\pm$0.73}               & 68.08{\tiny$\pm$1.41}                      & 71.42{\tiny$\pm$0.98}                    & \underline{72.04}{\tiny$\pm$1.09}                     & \textbf{74.71}{\tiny$\pm$1.41}                              \\
			\hline
			\multirow{5}{*}{10000 $\times$ 8 $\times$ 7}                                    & In-1 $\sim$ In-7                  & 42.30{\tiny$\pm$0.14}                   & \underline{94.76}{\tiny$\pm$1.82}                   & 93.76{\tiny$\pm$0.31}               & 39.65{\tiny$\pm$0.09}                      & 87.97{\tiny$\pm$0.92}                   & 89.05{\tiny$\pm$0.30}                   & 65.25{\tiny$\pm$0.41}               & 91.15{\tiny$\pm$1.48}                      & 89.44{\tiny$\pm$0.35}                    & 88.80{\tiny$\pm$0.18}                     & \textbf{95.11}{\tiny$\pm$0.23}                              \\
			& In-8                                  & 24.53{\tiny$\pm$0.42}                   & 55.93{\tiny$\pm$3.28}                   & 65.26{\tiny$\pm$0.94}               & 24.17{\tiny$\pm$0.19}                      & 61.36{\tiny$\pm$1.38}                   & 62.57{\tiny$\pm$1.13}                   & 41.12{\tiny$\pm$1.08}               & 59.78{\tiny$\pm$1.90}                      & \underline{66.23}{\tiny$\pm$0.93}                    & 62.66{\tiny$\pm$2.35}                     & \textbf{68.13}{\tiny$\pm$0.85}                              \\
			& In-9                                  & 28.50{\tiny$\pm$0.44}                   & 49.76{\tiny$\pm$5.64}                   & \textbf{60.78}{\tiny$\pm$1.15}               & 30.11{\tiny$\pm$0.14}                      &59.25{\tiny$\pm$1.37}                   & 57.19{\tiny$\pm$1.42}                   & 45.61{\tiny$\pm$0.40}               & 51.03{\tiny$\pm$1.53}                      & 58.43{\tiny$\pm$0.58}                    & 58.62{\tiny$\pm$1.42}                     & \underline{60.71}{\tiny$\pm$1.81}                              \\
			& Out-1                                 & 22.21{\tiny$\pm$0.24}                   & 38.65{\tiny$\pm$2.78}                   & \underline{49.24}{\tiny$\pm$1.23}               & 20.57{\tiny$\pm$0.22}                      &42.73{\tiny$\pm$1.29}                   & 42.65{\tiny$\pm$1.07}                   & 38.78{\tiny$\pm$0.49}               & 42.81{\tiny$\pm$2.22}                      & 45.51{\tiny$\pm$0.73}                    & 47.58{\tiny$\pm$0.72}                     & \textbf{49.69}{\tiny$\pm$0.94}                              \\
			& Out-2                                 & 14.00{\tiny$\pm$0.12}                   & 59.54{\tiny$\pm$3.45}                   & 66.02{\tiny$\pm$1.40}               & 13.95{\tiny$\pm$0.09}                      &61.80{\tiny$\pm$1.90}                   & 61.73{\tiny$\pm$1.20}                   & 35.56{\tiny$\pm$0.63}               & 61.84{\tiny$\pm$1.78}                      & \textbf{70.90}{\tiny$\pm$0.57}                    & 62.23{\tiny$\pm$1.17}                     & \underline{69.47}{\tiny$\pm$1.25}                 \\            
			\bottomrule
			\multicolumn{10}{l}{\tiny $\bullet$ The item in ``Number of samples for training'' is (Number of Samples) $\times$ (Number of Emitters) $\times$ (Number of Sessions).}\\ 
			\multicolumn{11}{l}{\tiny $\bullet$ The best outcome among all competitive methods is denoted in \textbf{bold}, while the second best performance is indicated with \underline{underline}.}
		\end{tabular}
	}
\end{table}

All models are implemented using PyTorch and trained on an NVIDIA Tesla V100 GPU.   The hyperparameter settings for experiments are detailed in Table~\ref{table:hyperparameters}.

In this experiment, we focus on evaluating the robustness of various models under different channel conditions. Channel variations, caused by differences in scenarios or the timing of data collection across sessions, can significantly affect the received signals and result in a sharp decline in model performance. A robust model should be less affected by these channel variations and better able to adapt to new channels and scenarios after being trained on a dataset collected in known environments.
we train all of the methods with the training sets from seven indoor sessions (In-1 to In-7), and then evaluate their performance using the test datasets from these seven sessions, as well as from In-8, In-9, and two outdoor sessions, Out-1 and Out-2. Additionally, we assessed how these methods perform with varying amounts of training data by using different number of samples: 40,000, 30,000, 20,000, and 10,000 samples per emitter from each session's training set.

The evaluation results of the experiment, presented in Table~\ref{table:compare-to-others}, demonstrate the excellent performance of SinFormer, particularly when combined with MAE-AC general feature learning. For the testing on indoor sessions In-1 to In-7, SinFormer with MAE-AC achieved an impressive accuracy of 99.6\% when sufficient data is available (40,000 samples per emitter per session). Even more notably, it maintains an accuracy of over 95\% across all training data sizes, significantly outperforming most of the methods.

\begin{figure}[t] \centering
	\captionsetup{font={small}}
	\subfloat[GRU-FCN]{\includegraphics[width=0.45\columnwidth]{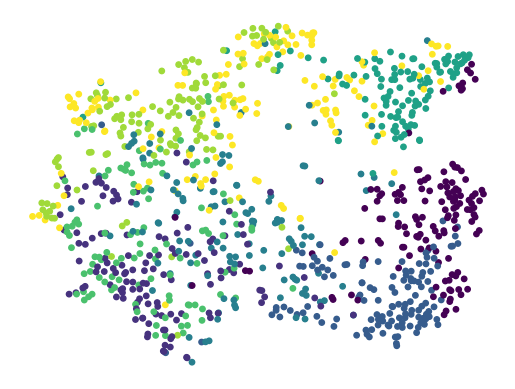}  
		\label{fig:GRUFCN_TSNE}  } \quad
	\subfloat[ViT]{\includegraphics[width=0.45\columnwidth]{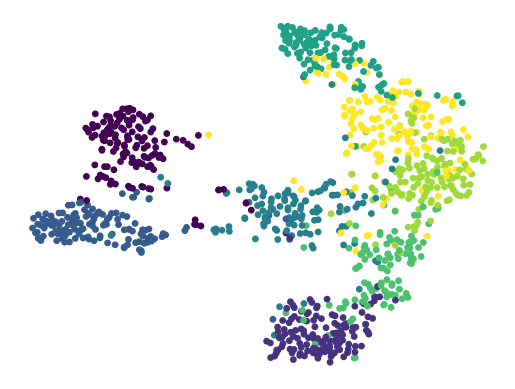}
		\label{fig:ViT_TSNE}  } \quad
	\subfloat[GLFormer]{\includegraphics[width=0.45\columnwidth]{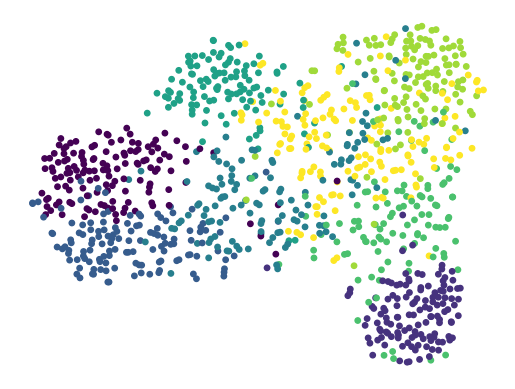}
		\label{fig:GLFormer_TSNE} } \quad
	\subfloat[SinFormer + MAE-AC]{\includegraphics[width=0.45\columnwidth]{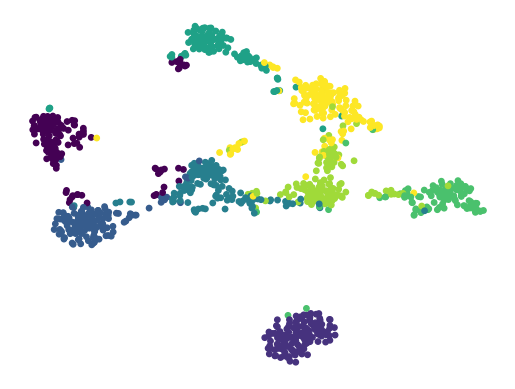}
		\label{fig:SinFormer_TSNE} }
	\caption{The t-SNE visualizations of extracted unseen session Out-2 signal feature distribution for different methods trained on In-1 to In-7 sessions, when training data is sufficient. Features
		with the same emitter label are plotted in the same color.}
	\label{fig:tsne}  
\end{figure}

The results on the In-8, In-9, Out-1, and Out-2 sessions demonstrate the models' ability to generalize to new environments and time sessions. As expected, all models show some decline in performance on these unseen sessions due to differences in channel temporal variations. However, SinFormer with two-stage training (MAE-AC) consistently outperforms other models, demonstrating its robustness and adaptability.
  On the In-8 test set, SinFormer surpasses the second-best method by 2.1\%. Similarly, on the In-9 test set, SinFormer achieves a 1.3\% higher accuracy. The performance gap widens on the outdoor test sets, with SinFormer outperforming the next best method by 2.2\% on Out-1 and an impressive 6.0\% on Out-2.
In addition, the t-distributed Stochastic Neighbor Embedding (t-SNE)~\cite{van2008visualizing} visualizations in Figure~\ref{fig:tsne} illustrate the feature distributions of the unseen Out-2 session signals, as extracted by different methods. The SinFormer with MAE-AC with two-stage training (MAE-AC) produces the most distinct and well-separated clusters among all the methods. The clear separation of features with the same emitter label shows that this method is highly effective at distinguishing between different emitters, even in an unseen session, demonstrating superior generalization and robustness.

The experiment also evaluates the impact of number of training samples on model performance. Despite the reduction in training data, SinFormer with MAE-AC consistently outperforms the other methods, maintaining a relatively high accuracy even with the small dataset. This results show the effectiveness of the two-stage training in enhancing SinFormer’s generalization ability.

\begin{table}[t] \scriptsize
	\renewcommand\arraystretch{1.3}
	\centering
	\caption{\label{table:results_ablation}Classification accuracy (\%) of experiments for the ablation study}
	\resizebox{1.0\columnwidth}{!}{
	\begin{tabular}{c c c |c c c c c | c c c c c}
		\toprule
		\multicolumn{3}{c|}{\textbf{Components}} & \multicolumn{5}{c|}{40000$\times$8$\times$7 training samples} & \multicolumn{5}{c}{20000$\times$8$\times$7 training samples} \\
		\hline
		\textbf{Multi-scale} & \textbf{SSAT} & \textbf{Two-stage training}  & In-1$\sim$In-7 & In-8 & In-9 & Out-1 & Out-2 & In-1$\sim$In-7 & In-8 & In-9 & Out-1 & Out-2\\ 
		\hline
		$\times$ & $\times$ & $\times$ & \textbf{99.71} & 65.91 & 61.02 & 52.84 & 70.97 & \textbf{98.15} & 65.12 & 62.75 & 51.26 & 70.04\\
		$\times$ & $\checkmark$ & $\times$ & 99.75 & 66.94 & 60.55 & 55.64 & 75.57 & 98.24 & 67.05 & 60.76 & 51.95& 71.72\\
		$\checkmark$ & $\times$ & $\times$ & 99.59 & 66.52 & 62.23 & 53.27 & 74.84 & 96.25 & 65.46 & 62.21 & 50.62 & 70.33\\
		$\checkmark$ & $\checkmark$ & $\times$ & 99.36 & 67.16 & \textbf{63.97} & \textbf{57.08} & \textbf{82.39} & 96.33 & 69.58 & 61.18 & 52.69 & 72.04\\
		$\checkmark$ & $\times$ & MAE-AC & 99.59 & 68.81 & 63.15 & 52.38 & 79.90 & 96.81 & 67.95 & 62.55 & 49.00 & 72.61\\
		$\checkmark$ & $\checkmark$ & MAE & 99.50 & 68.39 & 62.40 & 55.94 & 80.67 & 97.31 & 69.50 & 62.31 & 53.66 & 73.64\\
		$\checkmark$ & $\checkmark$ & MAE-AC & 99.60 & \textbf{69.10} & 63.13 & 55.32 & 82.02 & 97.38 & \textbf{70.07} & \textbf{62.88} & \textbf{53.91} & \textbf{74.71}\\
		\bottomrule
	\end{tabular}
}
\end{table}

\subsection{Ablation study}

In the ablation study, we conducted experiments to evaluate the impact of each component in the proposed method, focusing on multi-scale, SSAT, and the two-stage training approach. The hyperparameter settings used in these experiments are consistent with those in Section~\ref{sec:results}. The results are shown in Table~\ref{table:results_ablation}.

From Table~\ref{table:results_ablation}, it is observed that multi-scale feature extraction clearly enhances model performance. When comparing rows that multi-scale processing is utilized (the second and the fourth rows) against those that it is not (the first and the third rows), there is a noticeable improvement in accuracy across most test scenarios. For instance, with 40,000 samples per class per session for training, the accuracy on the Out-2 test set improves from 70.97\% without multi-scale processing to 74.84\% with it.

The utilization of SSAT also contribute to the better performance of the proposed model. When SSAT is used without two-stage training (the second and fourth rows in Table~\ref{table:results_ablation}), there is an increase in the identification accuracy, particularly on the unseen sessions. When data is sufficient and multi-scale feature extraction is employed, incorporating SSAT results in performance improvements across all unseen sessions compared to not using SSAT, with increases of 1.4\%, 1.7\%, 3.8\%, and 7.5\% on the In-8, In-9, Out-1, and Out-2 sessions, respectively.

From Table~\ref{table:results_ablation}, we have that two-stage training further boosts the model performance. In the case of 20,000 samples per class per session, using MAE-AC leads to performance improvements across all sessions compared to not using it, with increases of 1.0\%, 0.5\%, 1.7\%, 1.2\%, and 1.9\% on the In-1 to In-7, In-8, In-9, Out-1, and Out-2 sessions, respectively. Additionally, using MAE-AC during the general feature learning stage shows further improvements over using MAE, demonstrating that MAE-AC is more effective than MAE due to the utilization of the auxiliary classification task.

The results of the ablation study indicate that when training data is sufficient, multi-scale feature extraction and SSAT are the key contributors to performance improvement. However, as the amount of training data decreases, the two-stage training framework becomes increasingly crucial for maintaining and enhancing the model's performance, as this framework allows the model to learn generalized features during general feature learning stage, which can be fine-tuned to adapt to specific tasks with limited labeled data.

\begin{figure} \centering
		\subfloat[GRU-FCN]{\includegraphics[width=0.30\columnwidth]{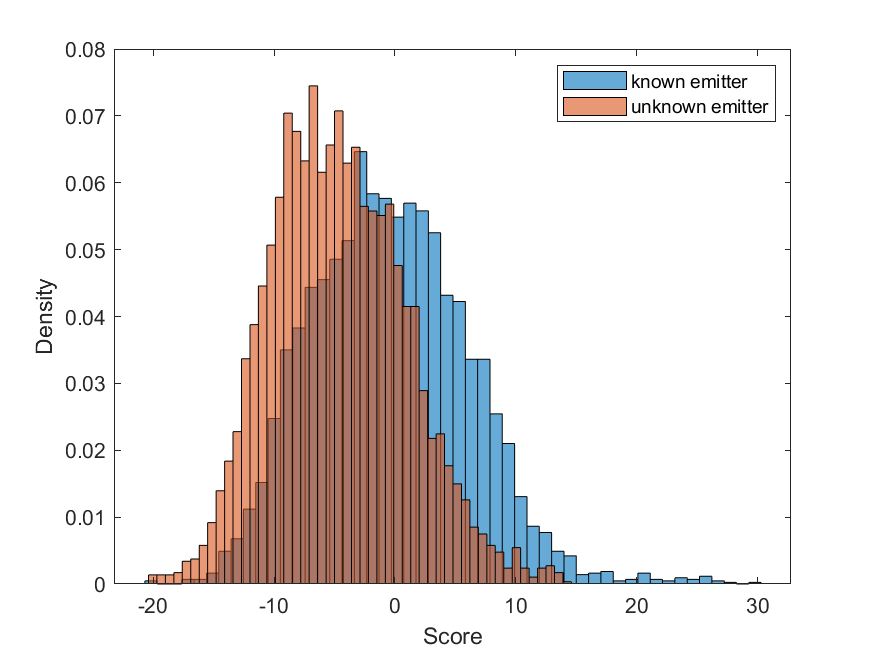}  
		\label{fig:ood_GRUFCN}  } \quad
	\subfloat[ResNext1D]{\includegraphics[width=0.30\columnwidth]{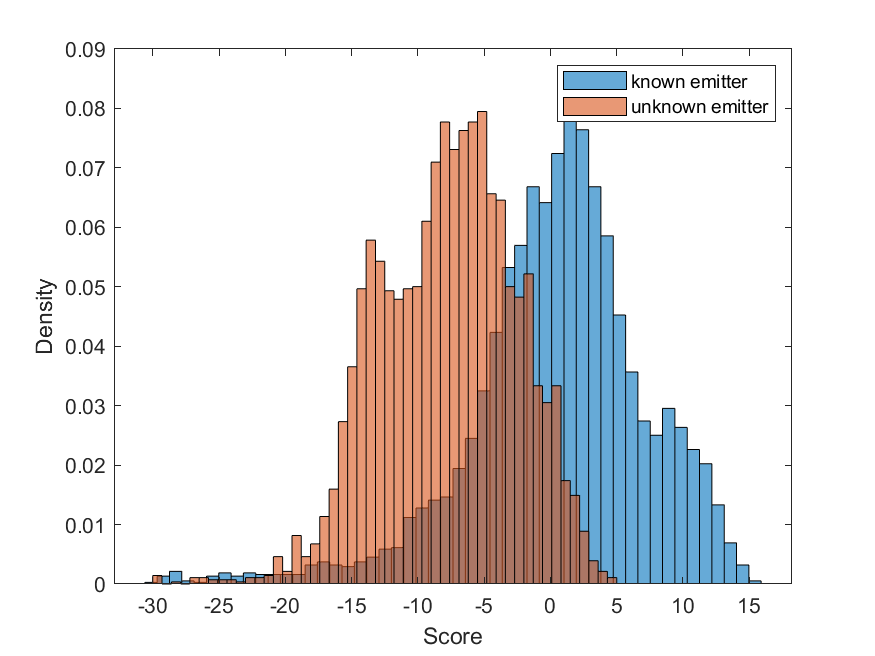}  
		\label{fig:ood_resnext1d}  } \quad
	\subfloat[ViT]{\includegraphics[width=0.30\columnwidth]{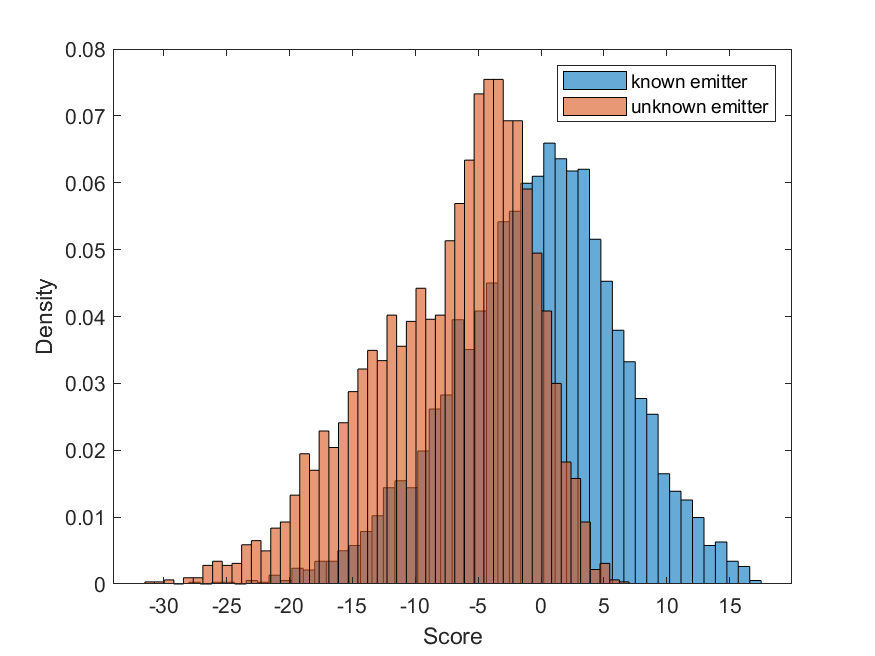}
		\label{fig:ood_ViT}  } \quad
	\subfloat[PatchMixer]{\includegraphics[width=0.30\columnwidth]{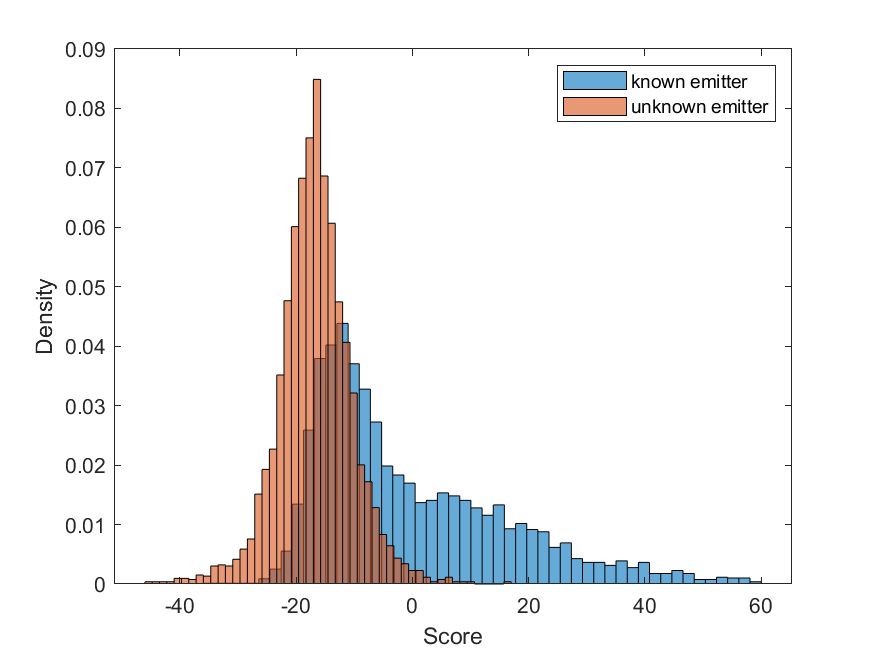}
		\label{fig:ood_patchmixer}  } \quad	
	\subfloat[PVT]{\includegraphics[width=0.30\columnwidth]{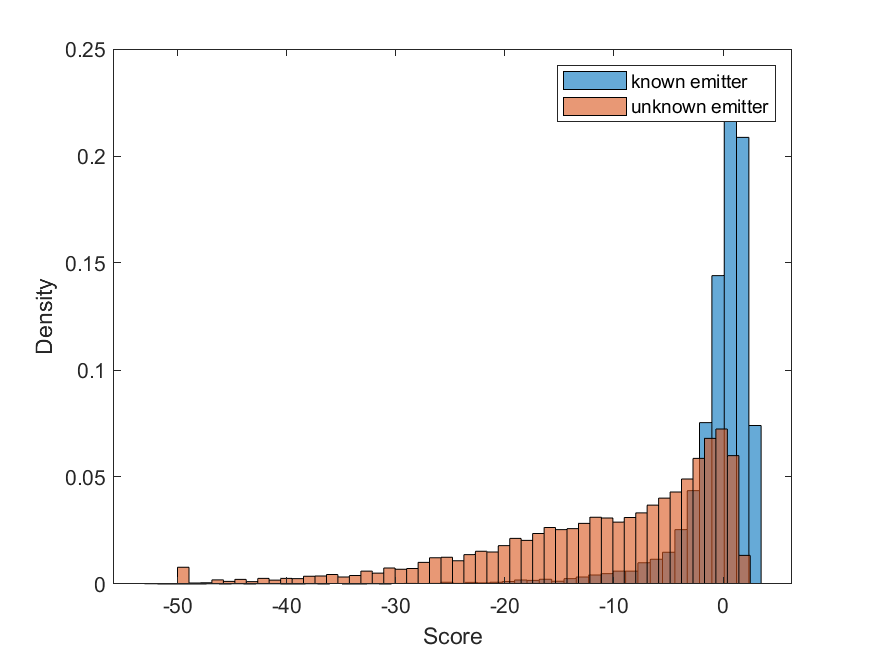}
		\label{fig:ood_PVT}  } \quad
	\subfloat[DeiT]{\includegraphics[width=0.30\columnwidth]{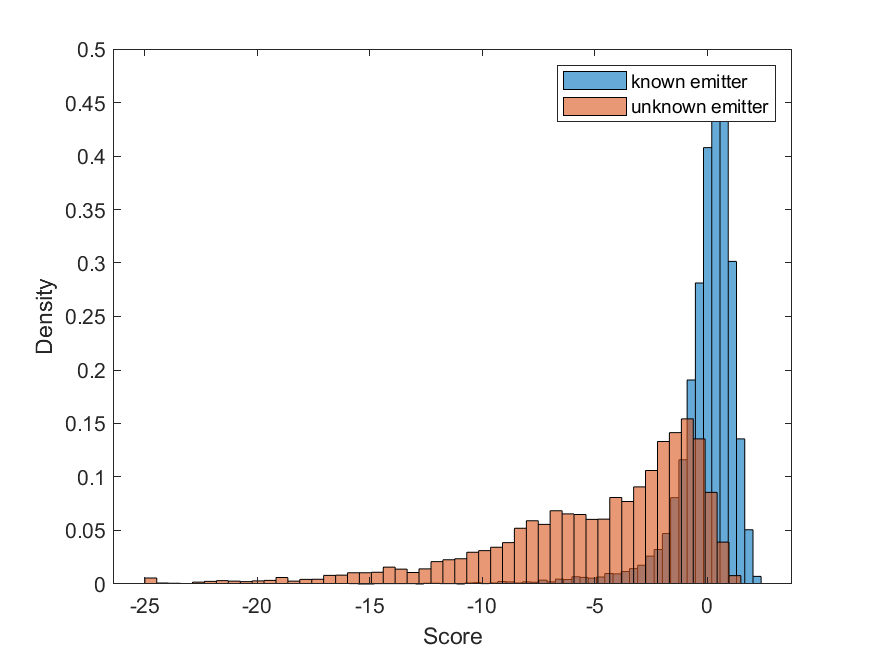}
		\label{fig:ood_DeiT}  } \quad	
	\subfloat[Vision Mamba]{\includegraphics[width=0.30\columnwidth]{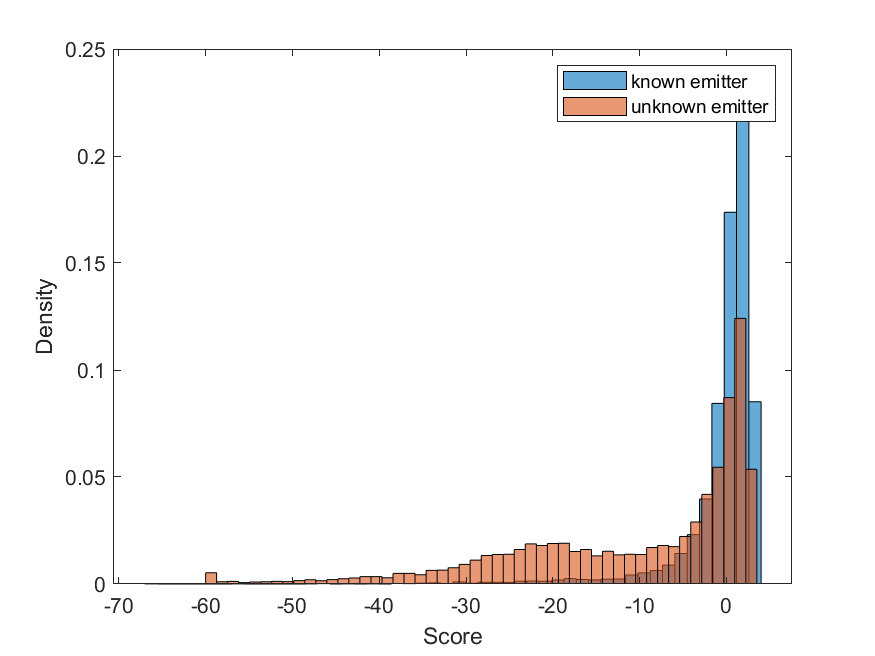}
		\label{fig:ood_vim}  } \quad	
	\subfloat[ASA]{\includegraphics[width=0.30\columnwidth]{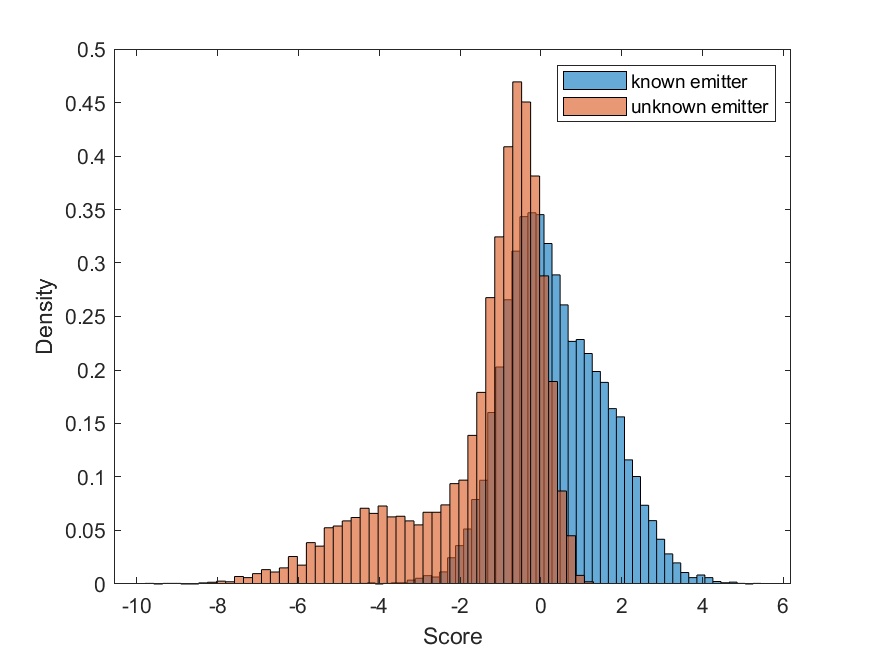}
		\label{fig:ood_asa}  } \quad		
	\subfloat[GLFormer]{\includegraphics[width=0.30\columnwidth]{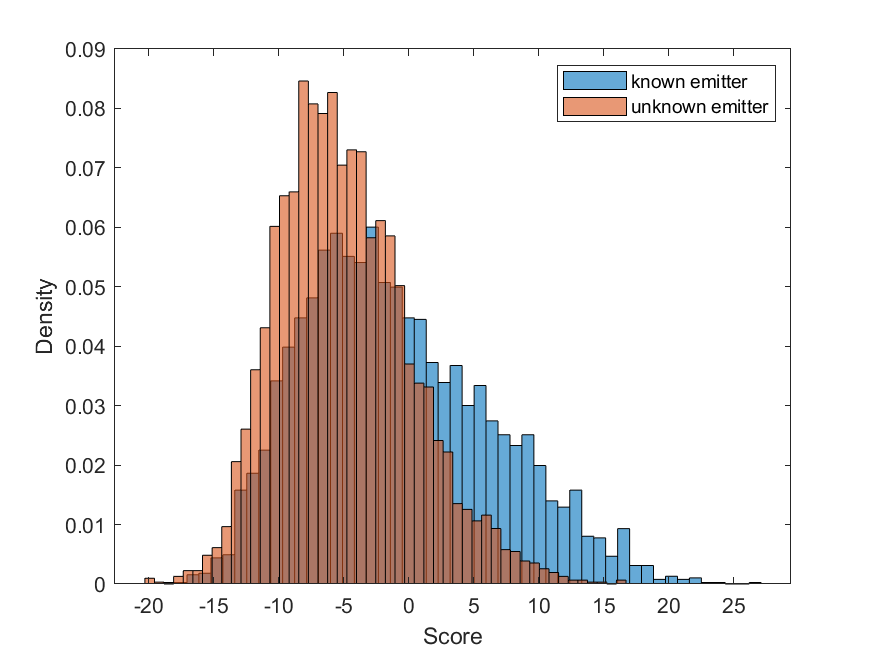}
		\label{fig:ood_GLFormer} } \quad
	\subfloat[SinFormer + MAE-AC]{\includegraphics[width=0.30\columnwidth]{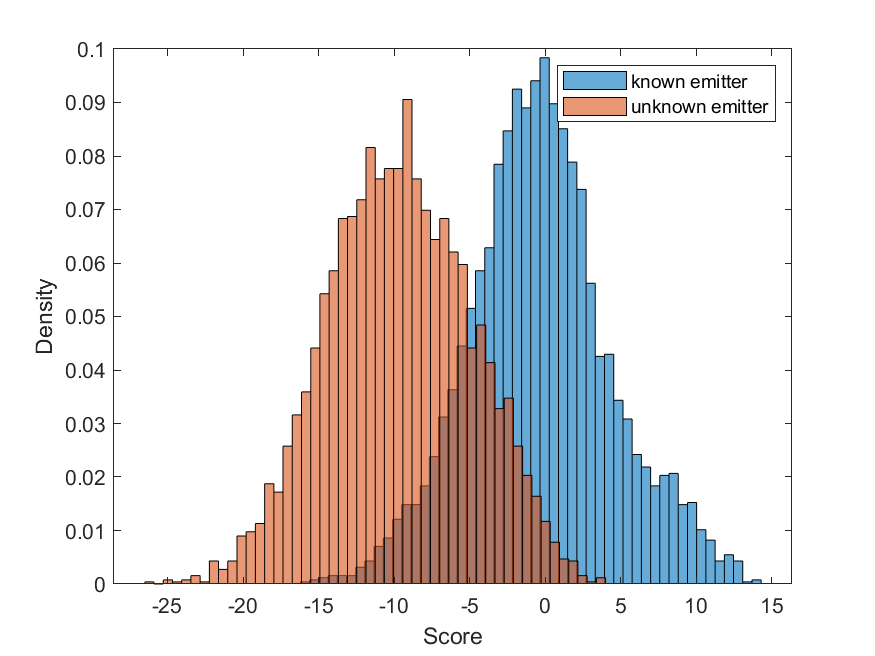}
		\label{fig:ood_SinFormer} }
	\caption{Histograms of known and unknown emitter scores for each method using the open-set recognition approach ViM.}
	\label{fig:ood}  
\end{figure}

\begin{table}[t] \scriptsize
	\renewcommand\arraystretch{1.3}
	\centering
	\caption{\label{table:results_openset}Results (\%) of experiments for open-set recognition}
	\begin{tabular}{c | c | c }
		\toprule
		\multirow{2}{*}{\textbf{Method}} & \multicolumn{2}{c}{\textbf{Metric}} \\
		\cline{2-3}
		~ & AUROC $\uparrow$ & FPR95 $\downarrow$ \\
		\hline
		GRU-FCN & 69.10 & 84.00 \\
		ResNext & 83.58 & 93.88 \\
		ViT &     78.31 & 75.24 \\
		PatchMixer & 85.02 & 62.10 \\
		  PVT &   87.19 &   47.65 \\
		  DeiT &   90.72 &   39.74 \\
		  Vision Mamba &   75.37 &   56.96 \\
		  ASA &   81.10 &   66.99 \\
		GLFormer & 66.23 & 91.19 \\
		SinFormer + MAE-AC & \textbf{92.04} & \textbf{35.60} \\
		\bottomrule
	\end{tabular}
\end{table}

\subsection{Results of open-set recognition}

\begin{figure}[t]
	\centerline{\includegraphics[width=0.9\columnwidth]{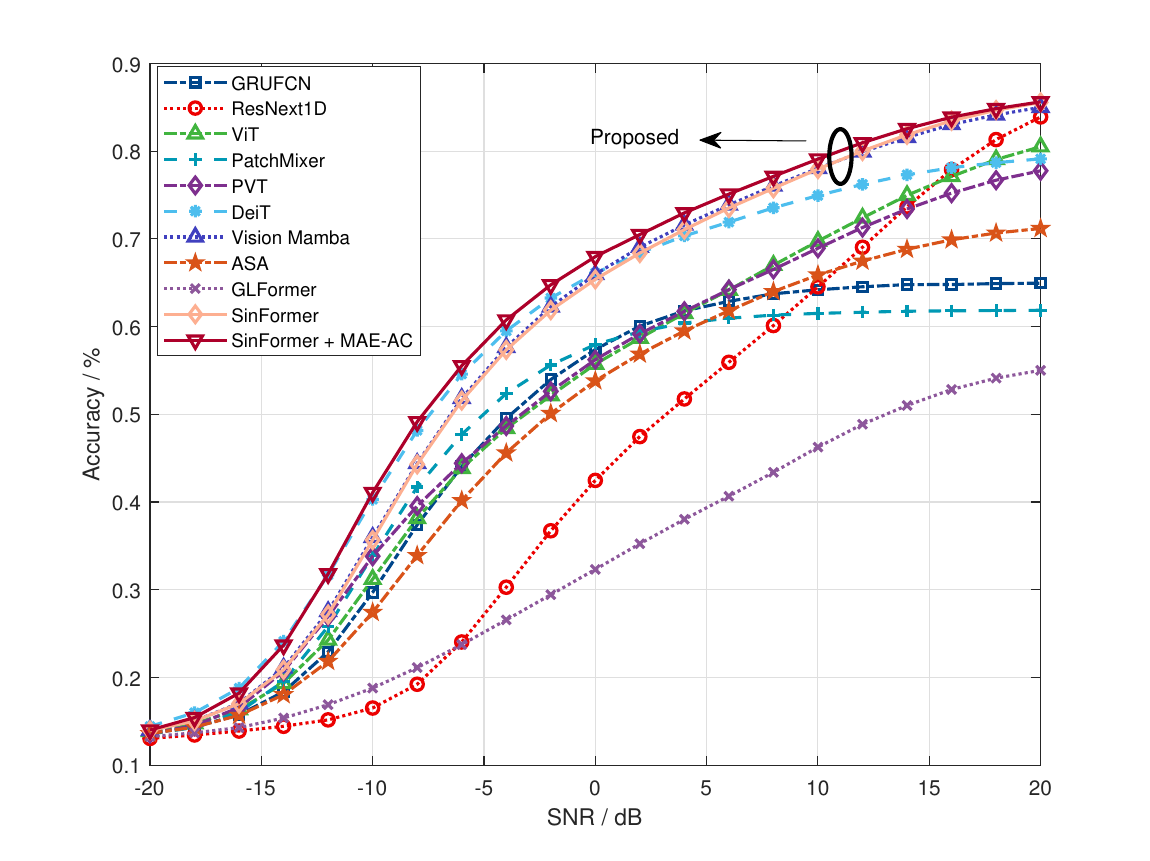}}
	\caption{Performance results of various methods across SNR ranges from -20 dB to 20 dB.}
	\label{fig:snr}  
\end{figure}

The ability to recognize signals from unknown emitters is a fundamental capability of any RFFI system, known as open-set recognition (OSR). In this experiment, we evaluated the OSR performance of various methods mentioned in Section~\ref{sec:results}. Specifically, we selecte 6 out of 8 transmitters as known emitters and the remaining 2 as unknown. The methods are trained using the data from the known emitters in sessions In-1 to In-7. To perform open-set recognition, we apply a post-processing OSR method, called Virtual-logit Matching (ViM)~\cite{wang2022vim}, for unknown emitter identification. The performance is assessed using Area Under the Receiver Operating Characteristic curve (AUROC) (higher is better) and False Positive Rate at 95\% True Positive Rate (FPR95) (lower is better).

The experimental results, as shown in Table~\ref{table:results_openset}, show that SinFormer with two-stage training (MAE-AC) achieves the best performance on both the AUROC and FPR95 metrics, significantly outperforming the other methods. Figure~\ref{fig:ood} presents the histograms of scores for known and unknown emitters obtained using ViM across various methods. It further illustrates each method's ability to distinguish between known and unknown emitters. And SinFormer with two-stage training (MAE-AC) exhibiting the highest level of discrimination. This result highlights the proposed method's powerful feature extraction capabilities and its exceptional ability to generalize to unseen data.

\subsection{Results of experiments with different SNRs} \label{sec:snr}

Noise resistance is a crucial indicator of a method's robustness. In this experiment, we evaluate the performance of various methods across different SNRs. We treat the collected signals as noise-free and then introduce the Gaussian noise at varying levels to the signals from the In-1 to In-7 sessions, generating SNRs ranging from -20 dB to 20 dB. These noisy signals are used for training, and the performance of each method is subsequently assessed across the In-1 to In-7 sessions at each SNR level.


The results are shown in Figure~\ref{fig:snr}. It demonstrates that SinFormer with two-stage training (MAE-AC) is the most robust method, achieving the highest accuracy across all SNR levels. This highlights its ability to maintain strong performance even in highly noisy environments. Without two-stage training, the baseline SinFormer achieves comparable performance to Vision Mamba, a recently proposed strong general-purpose model. However, with the incorporation of MAE-AC, SinFormer significantly surpasses Vision Mamba, particularly under low-SNR conditions. These results suggest that while the base SinFormer is already competitive, the two-stage training strategy enables the model to extract more noise-invariant and generalizable features, leading to consistent gains across varying SNR levels.

\subsection{Results of robustness in other practical scenarios}
To further evaluate the robustness of the model, we used the same trained models as in the SNR robustness experiments in Section~\ref{sec:snr}, without additional fine-tuning or retraining.
We consider the following two types of interference:

\textbf{Narrowband Interference}: In our simulation, the entire signal frequency band is divided into 256 sub-bands. For each data sample, two sub-bands are randomly selected and corrupted with narrowband interference. The selection is fixed within each individual sample but randomized across different samples. We evaluated the model performance across different signal-to-interference ratios (SIRs) to assess its sensitivity to frequency-localized disturbances, which may occur due to nearby interfering devices or partial-band jamming.

\textbf{Multipath Effects:}
To simulate multipath propagation, we followed the channel model described in~\cite{Fadul2019rf}, where the received signal consists of two paths with a fixed delay of 150 ns. The first path is considered the desired signal, and the second path is treated as an interfering reflection. By varying the SIR between these two paths, we evaluated how well the model tolerates multi-path effects.

\begin{figure}[t] \centering
	\subfloat[Narrowband interference]{\includegraphics[width=0.47\columnwidth]{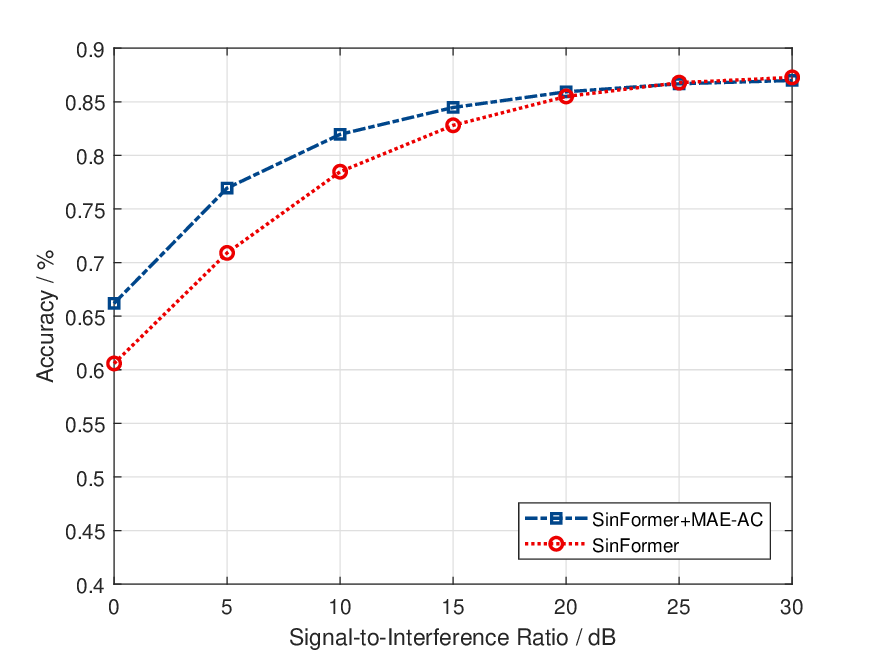}  } \quad
	\subfloat[Multi-path effects]{\includegraphics[width=0.47\columnwidth]{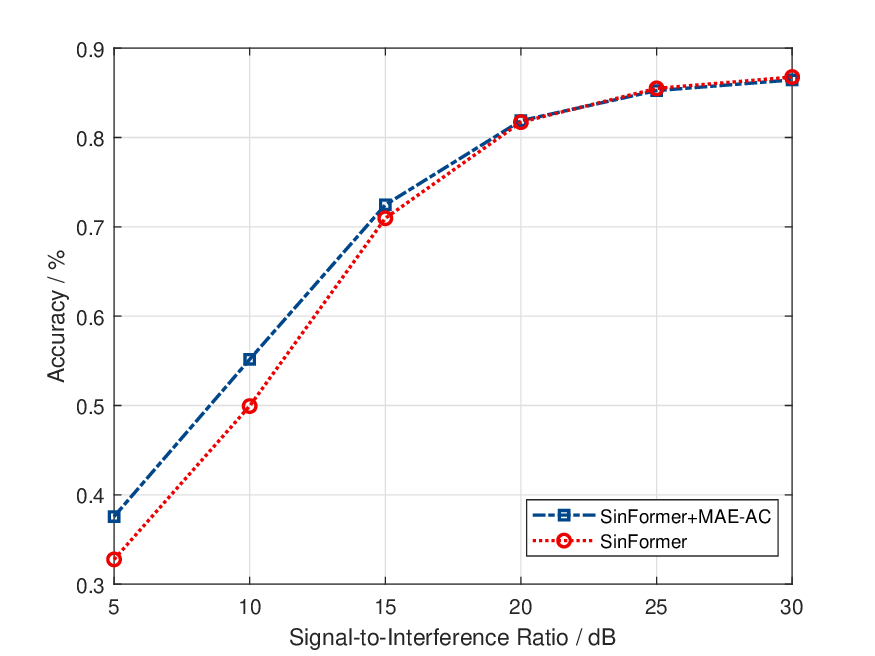}} \quad
	\caption{\label{fig:interference}Accuracy vs. SIR in other practical scenarios.} 
\end{figure}

Figure~\ref{fig:interference} demonstrates the robustness of the proposed method under narrowband interference and multipath conditions. As the SIR decreases, both variants of SinFormer maintain relatively high accuracy, indicating strong resilience to practical interference. In particular, the model exhibits greater robustness against narrowband interference compared to multipath effects. This may be attributed to the fact that narrowband interference only partially disturbs the signal spectrum, allowing the model to effectively suppress it by leveraging multi-scale features. In contrast, multipath interference introduces complex time-domain distortions, which are generally more challenging to resolve.

\begin{figure}[t] \centering
	\subfloat[Hyper-parameters $\alpha, \beta$]{\includegraphics[width=0.47\columnwidth]{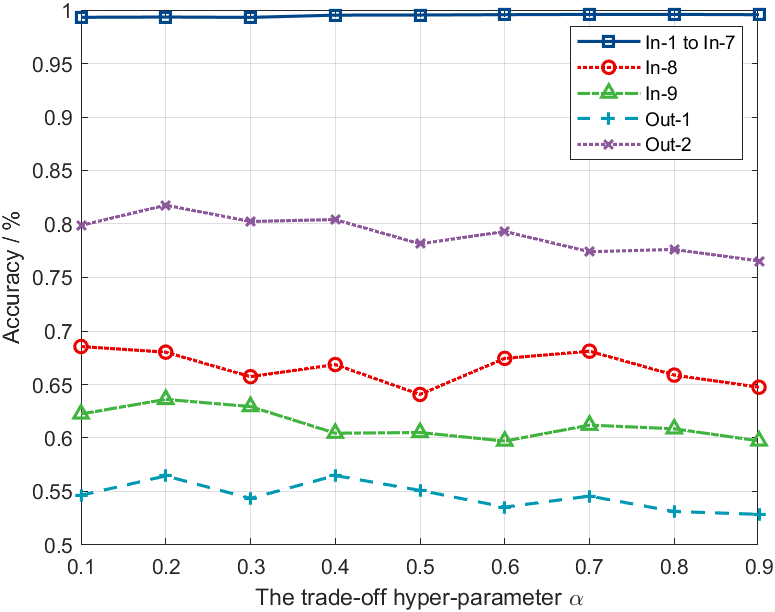}  
		\label{fig:alpha}  } \quad
	\subfloat[The number of blocks $L$]{\includegraphics[width=0.47\columnwidth]{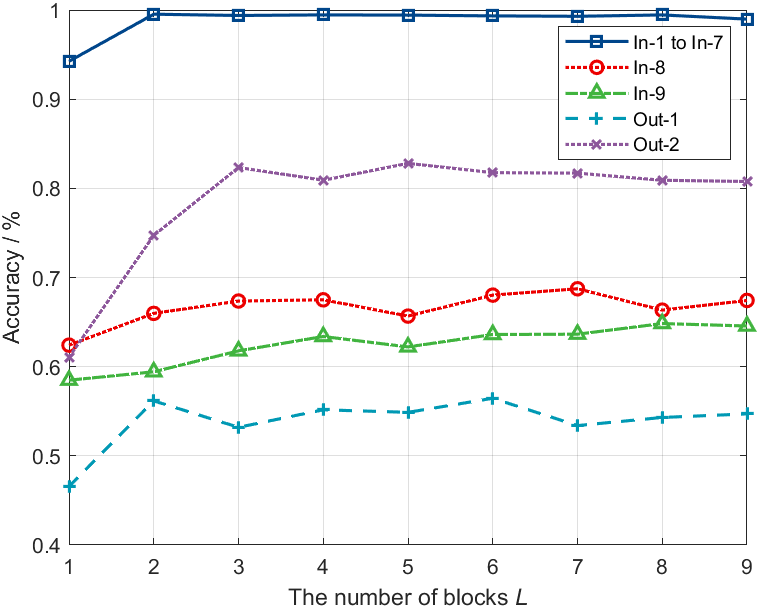}
		\label{fig:L}  } \quad
	\caption{Impact of hyper-parameter variations on results on In-1 to In-7, In-8, In-9, Out-1, and Out-2 Sessions}
	\label{fig:hyperparameter}  
\end{figure}

\subsection{Results of hyper-parameter sensitivity}

\subsubsection{The trade-off hyper-parameters $\alpha, \beta$}

For simplicity in selecting hyperparameters, we set $\beta$ as $1-\alpha$ and varied $\alpha$ from 0.1 to 0.9. The experimental results, shown in Figure~\ref{fig:hyperparameter}(a), indicate that accuracy across the sessions remains relatively unchanged as $\alpha$ varies. However, when $\alpha$ approaches 1, there is a slight decrease in accuracy on the unseen sessions. This decline may be due to SSAT acting as a regularizer that adjusts the gradient direction, and a larger $\beta$ ensures that SSAT's corrective influence is effective.

\subsubsection{The number of blocks $L$}

We also conducted hyperparameter experiments to evaluate the impact of the number of blocks $L$ in SinFormer, which directly influences the numbers of the parameters. The value of $L$ is varied from 1 to 9. As shown in Figure~\ref{fig:hyperparameter}(b), when $L$ is small, increasing the number of blocks significantly enhances the model’s performance, leading to higher accuracy across all sessions. However, once $L$ exceeds a certain threshold ($L \geq 4$), the accuracy shows no substantial improvement with additional blocks. This indicates that beyond a certain level of complexity, further increasing the number of blocks provides diminishing returns, as the model has likely captured the essential features for optimal performance.

\subsection{Results of scalability}
To evaluate the scalability of the proposed SinFormer model, we conducted additional experiments on the ManyTx subset of the publicly available WiSig dataset~\cite{hanna2022wisig}. WiSig is a large-scale WiFi RF fingerprinting dataset comprising over 10 million packets collected from 174 commercial WiFi transmitters and 41 USRP receivers over a one-month period.

The ManyTx subset of Wisig dataset focuses on transmitter diversity and includes data from 150 transmitters and 18 receivers, recorded over 4 days, with 50 signal samples per transmitter-receiver pair per day. This results in a total of 540,000 samples. To simulate different dataset scales, we progressively selected subsets with 10, 30, 50, 70, 90, 110, 130, and 150 transmitters. For each transmitter, 80\% of the samples per receiver per day were used for training, and the remaining 20\% were used for testing.

\begin{figure}[!t]
	\centerline{\resizebox{.7\textwidth}{!}{\includegraphics{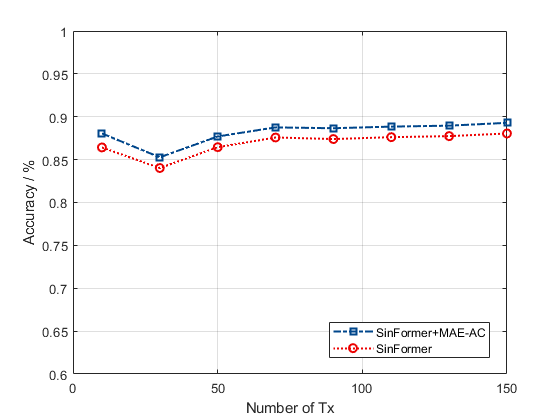}}
	}   \caption{The scalability of proposed method} \label{fig:scalability}
\end{figure}

As shown in Figure~\ref{fig:scalability}, the proposed method exhibits good scalability across varying numbers of transmitters. SinFormer with two-stage training consistently achieves high accuracy, reaching up to 89.5\% with 150 devices, while the baseline SinFormer also maintains stable performance above 88\%. These results confirm that the proposed model generalizes well to large-scale RFFI tasks and demonstrates strong scalability.

\section{Conclusions and Discussions}\label{sec:conclusion}

In this paper, we introduced a novel deep learning-based framework for RFFI, designed to enhance the accuracy, reliability, and robustness of device identification in complex and dynamic environments. 
The proposed SinFormer model incorporates a multi-scale self-attention mechanism to effectively capture diverse features of radio signals, enabling precise identification. Furthermore, the two-stage training strategy improves the model's adaptability to varying conditions. 
Experimental results on real-world datasets demonstrate that our approach outperforms existing methods, offering a more robust and accurate solution for device identification. This framework establishes a new benchmark for future research in RFFI and holds promise for enhancing the security and reliability of wireless and IoT systems.

Although SinFormer achieves strong performance, its current design extracts multi-scale features uniformly across all predefined scales, regardless of the specific characteristics of the input data. This dense extraction may lead to redundant representations or unnecessary computational overhead, especially for datasets that predominantly require features at a particular scale. To address this limitation, future work will explore the integration of a Mixture-of-Experts (MoE) mechanism into the multi-scale extraction process. By dynamically selecting the most relevant scales based on the input, this design is expected to reduce redundancy, improve efficiency, and further enhance the model’s adaptability to diverse signal environments.

\bibliographystyle{elsarticle-num}
\bibliography{reference}
\end{document}